\documentclass[12pt]{article}
\usepackage{latexsym}
\usepackage{amssymb}
\usepackage{graphicx}

\newcommand{\be}{\begin{equation}} \newcommand{\ee}{\end{equation}}
\newcommand{\bea}{\begin{eqnarray}} \newcommand{\eea}{\end{eqnarray}}

\def\am{Center For Theoretical Physics, Department of Physics\\ Texas A\&M University,
 College Station,TX 77843-4242,
USA}

\def\address#1{\begin{center}{ \it #1} \end{center}}
\def\author#1{\begin{center}{ \sc #1} \end{center}}
\def\title#1{\begin{center} {\Large #1 } \end{center}}
\def\Journal#1#2#3#4{{#1} {\bf #2}, #3 (#4)}

\def\EJC{{\em Eur.Phys.J} C}
\def\JHEP{\em JHEP}

\def\NPB{{\em Nucl. Phys.} B}
\def\PLB{{\em Phys. Lett.}  B}
\def\PRL{\em Phys. Rev. Lett.}
\def\PRD{{\em Phys. Rev.} D}

\begin{document}

\begin{titlepage}

\title{Gravitational Forces in the Randall-Sundrum Model with a Scalar Stabilizing Field}
\author{R. Arnowitt and J. Dent}
\address{\am}

\begin{abstract}
We consider the problem of gravitational forces between point particles on the branes in a five dimensional (5D) Randall-Sundrum model with two branes (at $y_1$ and $y_2$) and $S^1/Z_2$ symmetry of the fifth dimension.  The matter on the branes is viewed as a perturbation on the vacuum metric and treated to linear order.  In previous work \cite{ad} it was seen that the trace of the transverse part of the 4D metric on the TeV brane, $f^T(y_2)$, contributed a Newtonian potential enhanced by $e^{2\beta y_2} \cong 10^{32}$ and thus produced gross disagreement with experiment.  In this work we include a scalar stabilizing field $\phi$ and solve the coupled Einstein and scalar equations to leading order for the case where $\phi_{0}^2/M_{5}^3$ is small and the vacuum field $\phi_{0}(y)$ is a decreasing function of $y$.  $f^T$ then grows a mass factor $e^{-\mu r}$ where however, $\mu$ is suppressed from its natural value, $\mathcal{O}(M_{Pl})$, by an exponential factor $e^{-(1+\lambda_b)\beta y_2}$, $\lambda_b > 0$.  Thus agreement with experiment depends on the interplay between the enhancing and decaying exponentials.  Current data eliminates a significant part of the parameter space, and the Randall-Sundrum model will be sensitive to any improvements on the tests of the Newtonian force law at smaller distances.  An example of coupling of the $\phi$ field to the Higgs field is examined and found to generally produce very small effects.
\end{abstract}

\end{titlepage}

\section{Introduction}

Higher dimensional models in particle physics with dimension D $>$ 4 have been the subject of much theoretical investigation over the past two decades.  Higher dimensional theory arises naturally in string/M-theory and is phenomenologically interesting as they offer the possibility of explaining fundamental features of nature that would not be possible in 4D theory.  The simplest phenomenology of this type is the 5D Randall-Sundrum model (RS1)\cite{rs,rs2} where the fifth dimension y is compactified with $S^1/Z_2$ symmetry so that one can think of space as bounded by two 4D orbifold planes (3-branes) at $y_1 = 0$ and $y_2 = \pi\rho$ with boundary conditions at $y_1$ and $y_2$ to enforce the $S^1/Z_2$ symmetry.  With no matter on the branes, the 5D Einstein equations have a vacuum solution which preserves 4D Poincaire invariance on the branes
\begin{equation}
ds^2 = e^{-2A(y)}\eta_{ij}dx^i dx^j + dy^2
\end{equation}
where
\begin{equation}
A(y) = \beta|y|\,\,\,;\,\,\, y_1 - \epsilon \leq y \leq y_2 - \epsilon\,\,\,;\,\,\,\epsilon > 0
\end{equation}
and $\eta_{ij}$ is the Lorentz metric.  Thus if all basic masses are naturally of Planck size and the physical world lives on the $y_2$ brane, such a structure offers a new way of understanding the gauge hierarchy (without undue fine tuning) not available in 4D theory. For example, consider a scalar field $\chi$ on the $y_2$ brane which we may treat as a perturbation on the vacuum state. The action has the form
\begin{eqnarray}
S = -\frac{1}{2}\int d^4x\sqrt{-g}(g^{ij}\partial_i\chi\partial_j\chi + m^2\chi^2)
\end{eqnarray}
where we use the notation $\mu, \nu$ = 0,1,2,3,5 and i, j = 0,1,2,3.  Letting $\chi'$ = $e^{-\beta y_2}\chi$, the theory then takes canonical form with a mass parameter
\begin{eqnarray}
\bar{m} = e^{-\beta y_2}m
\end{eqnarray}
and the observed mass on the $y_2$ brane would be of TeV size if  $e^{-\beta y_2}$ $\simeq$ $10^{-16}$ i.e. $\beta y_2$ $\simeq$ 35. Thus a Planck size mass travelling on
the $y_2$ brane has its mass effectively supressed by the strong 5D gravitational forces (much as an electron traveling in a solid has its mass modified by the electric fields there).
The question remains, however, as to whether the 5D theory will produce other additional phenomena that would violate known observations on the physical $y_2$ brane. Initial analysis examined whether the Friedmann-Robertson-Walker (FRW) cosmology on the $y_2$ brane could be achieved in this model \cite{binetruy,csaki,khoury,dewolfe,cline,cline2,enq,ellwanger,lukas,kobayashi,add}.  This was found to indeed be the case provided that in addition to gravity being in the 5D bulk, one must stabilize the vacuum metric, which is most easily accomplished by adding a scalar field in the bulk, $\phi(x^i,y)$\cite{gw}. Then both relativistic and non-relativistic matter could be accommodated in the cosmology\cite{cline,cline2,add}, the distance between the branes being governed by the density of non-relativistic matter \cite{add}.

A second question that has been examined is whether the 5D theory correctly reproduces the known gravitational forces between particles. Here we treat the matter on the branes as a perturbation to the vacuum metric:
\begin{equation}
ds^2 = e^{-2\beta y}(\eta_{ij}+ h_{ij})dx^i dx^j + h_{i5}dydx^i + (1 + h_{55})dy^2
\end{equation}
There is also a large literature on this subject \cite{lykken,tanaka,giddings,chung,dorca,deruelle,nojiri,callin,ad} Refs.[15-22] deal only with the gravitational forces on the $y_1$ = 0 ``Planck'' brane (where all the masses are of Planck size), and find normal Newtonian forces hold between particles on the Planck brane (along with negligibly small Kaluza-Klein corrections). In previous work \cite{ad}, we have examined in addition the physically relevant forces on the $y_2$ ``TeV'' brane, and unlike other discussions make sure that the coordinate conditions chosen do not lead to bent branes (so that the $S^1/Z_2$ boundary conditions can be correctly imposed). To see what occurs for this case, it is convenient to make a 4D ADM decomposition of $h_{ij}$ \cite{adm,adm2}
\begin{equation}
h_{ij} = h_{ij}^{TT} + h_{ij}^{T} + h_{i,j} + h_{j,i}
\end{equation}
where $h_{ij}^{TT}$ is transverse and traceless ($\partial^i h_{ij}^{TT} = 0 = h^{i\,TT}_{i}$)and $h^{T}_{ij}$ is transverse  with a non-zero trace $f^T (\partial^i h_{ij}^T = 0, h^{i\,T}_{i} = f^T$). One may write
\begin{equation}
h_{ij}^{T} = \frac{1}{3}\pi_{ij}f^T \,\,\,;\,\,\, \pi_{ij} \equiv \eta_{ij} - O_{ij}
\end{equation}
where
\begin{equation}
O_{ij} \equiv \frac{\partial_i \partial_j}{\Box^2}
\end{equation}
(In the above and following, four dimesional indices are raised and lowered with the Lorentz metric $\eta_{ij}$.) What was found in \cite{ad} was that $h_{ij}^{TT}$ gave rise to leading order to normal Newtonian forces between particles on the Planck or TeV branes. However $f^T$ gave a Newtonian contribution on the TeV brane that was enhanced by a factor of $e^{2\beta y_2} \simeq  10^{32}$, thus producing a gross disagreement with experiment.  

None of the analyses discussed above, \cite{lykken,tanaka,giddings,chung,dorca,deruelle,nojiri,callin,ad}, have included a scalar field $\phi_{0}(x^i,y)$ \cite{gw} to stabilize the vacuum metric.  Such a scalar field might produce a mass for the $f^T$ field, thus modifying it gravitational potential. In this paper we examine the effects of introducing such a stablizing contribution. It is not possible to solve the coupled Einstein scalar field equations in closed form, but an iterative solution can be obtained when $\phi_{0}^2/M_{5}^3$ is small (where $\phi_{0}(y)$ is the vacuum solution and $M_{5}$ is the 5D Planck mass) and $\phi_{0}(y)$ is a decreasing function. Within this framework we find that $f^T$ indeed grows a mass $\mu$ but the mass is exponentially suppressed.  The $f^T$ contribution then appears effectively massless over a distance $r \lesssim 1/\mu$ which can be anomalously large due to the exponential reduction of $\mu$.  Thus whether the theory is in agreement
  with current experimental tests of the Newtonian force law at small distances depends on the interplay of the amount of suppression of $\mu$ compared to the size of the enhancement factor $e^{2\beta y_2}$ of the amplitude of $f^T$, and in fact current experiment strongly constrains allowable scalar field models of this type.

In Sec. 2, we give the choice of coordinate conditions we use and write down the Einstein and scalar field equations. In Secs. 3 and 4 we state the expansion procedure we use to solve the equations, discuss the solutions of the Einstein equations and evaluate $f^T$ at $y_1$ and $y_2$ (which is what is needed to calulate the effect of $f^T$ on particles on the branes). Sec. 5 is devoted to the scalar field equations, and in Sec. 6 we calculate the leading effects to the fields for a case where a rigorous vacuum solution for $\phi_{0}(y)$ exists.  In Sec.7 we calculate the Newtonian forces on particles on the branes. Sec. 8 examines a $\phi^2\chi^2$ coupling where $\chi$ is a scalar field (e.g. the Higgs field).  Conclusions are given in Sec.9.  Appendix A shows that the results obtained in Sec.7 are valid for a general class of models where $\phi_{0}^2/M_{5}^3$ is small and $\phi_{0}(y)$ is a decreasing function of y.  Appendix B discusses previous analyses of gravitational forces in RS1, and examines why they did not obtain the results given here.

\section{Coordinate Conditions}

The action for our system has the form
\begin{eqnarray}\nonumber
S &=& \int d^5x \sqrt{-g^5}[-\frac{1}{2}M_{5}^3R - \Lambda] + \int d^5x\sqrt{-g^5}[-\frac{1}{2}\partial^\mu\phi\partial_\mu\phi - V(\phi)] \\&+& \sum_{\alpha}\int d^5x\sqrt{-g^4}[\mathcal{L}_{m_{\alpha}} - V_{\alpha}(\phi)]\delta(y-y_{\alpha})
\end{eqnarray}
where $M_{5}$ is the 5D Planck mass, $R$ is the 5D curvature scalar, $\mathcal{L}_{m_{\alpha}}$ is the Lagrangian for point particles on the $y_{\alpha}$ brane ($\alpha$ = 1,2), $\phi(x^{\mu})$ is the scalar field that stabilizes the vacuum metric, $V(\phi)$ is the bulk potential, and $V_{\alpha}$ are the brane potentials.  We write
\begin{equation}
\phi(x^\mu) = \phi_{0}(y) + \delta\phi(x^\mu)
\end{equation}
where $\phi_{0}(y)$ is the vacuum solution and $\delta\phi$ is the perturbation due to matter on the branes.  The vacuum equations read
\begin{eqnarray}
4A'^2 - A'' = -\frac{2}{3M^{3}_5}[\Lambda + V(\phi_0)] -\frac{1}{3M_{5}^3}\sum_{\alpha}\delta(y-y_{\alpha})V_{\alpha}(\phi_0)\\
4A'^2 -4A'' = -\frac{2}{3M^{3}_5}[\Lambda + V(\phi_0)] -\frac{1}{M_{5}^3}\phi_{0}'^2 -\frac{4}{3M_{5}^3}\sum_{\alpha}\delta(y-y_{\alpha})V_{\alpha}(\phi_0)\\
\phi_{0}'' -4A'\phi_{0}' - V'(\phi_0) - \sum_{\alpha}\delta(y-y_{\alpha})V_{\alpha}'(\phi_0) = 0
\end{eqnarray}
where $A' \equiv dA(y)/dy$, $A'' \equiv d^2A/dy^2$, etc., and $V'(\phi_0) \equiv dV/d\phi_0$, etc.

The bulk and brane potentials are arbitrary except that they must be fine tuned to cancel the effects of the bulk cosmological constant $\Lambda$ so that the net brane cosmological constant vanishes.  Most of the analysis can be done without specifying $V$ and $V_{\alpha}$, making use of the field equations Eqs.(11-13).  However, to estimate the size of effects, it is useful to have an explicit rigorous solution of Eqs.(11-13) and one has been given in \cite{dewolfe}.  Thus vacuum functions
\begin{eqnarray}
A(y) = \beta y + \frac{1}{12}\frac{\phi_{1}^2}{M_{5}^3}e^{-2by}\\
\phi_{0}(y) = \phi_1 e^{-by}
\end{eqnarray}
are the solutions of Eqs.(11-13) for the choice 
\begin{eqnarray}
V(\phi_{0}(y)) &=& (\frac{1}{2}b^2 + 2b\beta)\phi_{0}^2 - \frac{b^2}{6M_{5}^3}\phi_{0}^4\\\nonumber
V_{\alpha}(\phi_{0}(y)) &=& V_{\alpha}(\phi_{0}(y_{\alpha})) + (-1)^{\alpha}2b\phi_{0}(y_{\alpha})(\phi_{0}(y) - \phi_{0}(y_{\alpha})) \\&+& \gamma_{\alpha}\frac{1}{2}(\phi_{0}(y)-\phi_{0}(y_{\alpha}))^2
\end{eqnarray}
where $\Lambda = -6M_{5}^3\beta^2$ (the fine tuning of the cosmological constant),
\begin{eqnarray}
V_{\alpha}(\phi_{0}(y_{\alpha})) = (-1)^{\alpha + 1}[6M_{5}^3\beta - b\phi_{0}^2(y_{\alpha})]
\end{eqnarray}
and $\gamma_{\alpha}$ are arbitrary constants.  We see that the effect of the scalar field is to add a term to A(y) of size $\phi_{0}^2/M_{5}^3$.  Since naturalness implies that all masses should be of the same order and comparable to $M_{Pl}$ we will assume $b \approx \beta$, and $b > 0$.  The gauge hierarchy condition then requires $\beta y_2 \simeq 35$ \cite{dewolfe} so that the addition to $A(y)$ is a rapidly decreasing quantity.

We assume that matter on the branes represent a perturbation to the vacuum state and so we solve the full field equations to first order in $h_{\mu\nu}(x^{\alpha})$ and $\delta\phi(x^{\alpha})$.  We begin by reviewing the coordinate conditions we will use in the following analysis.  The general transformation 
\begin{equation}
x'^{\mu} + \xi^{\mu} = x^{\mu}
\end{equation}
that preserves the $S^1/Z_2$ symmetry with no brane bending is constrained by
\begin{equation}
\xi^5(x^i,y_1) = 0 = \xi^5(x^i,y_2)
\end{equation}
As discussed in \cite{ad}, one may use these to set $h_{5i}$ to zero, but in general it is not possible to have $h_{55}$ vanish without introducing brane bending.  We thus assume in the following that
\begin{equation}
h_{5i} = 0\,\,\,;\,\,\,i=0,1,2,3
\end{equation}
There still remains some gauge freedom.  Thus under a general transformation preserving Eqs.(20, 21), the components of the metric transform to first order as \cite{ad}
\begin{eqnarray}
\delta h_{55} = 2\xi^{5}_{,5}\\
\delta h_{ij}^{TT} = 0\,\,\,;\,\,\, \delta f^T = -6A'\xi^5\\
\delta h_{i}^T = e^{2A}\xi_{i}^T\\
\delta(\Box^2 h^L) = 2e^{2A}\Box^2 \xi^L + 2A'e^{-2A}(e^{2A}\xi^L)_{,5}
\end{eqnarray}
where we have decomposed $h_i$ and $\xi_i$ into transverse and longitudinal parts, e.g. $h_i = h_{i}^T + h^L_{,i}$ where $\partial^i h_{i}^T \equiv 0$.  Eq.(21) requires in addition that
\begin{eqnarray}
\xi_{i}^T = e^{-2A(y)}F_{i}^T(x^i)\\
(e^{2A}\xi^L)_{,5} = -e^{2A}\xi_5
\end{eqnarray}
where $F_{i}^T(x^i)$ is an arbitrary function independent of y.  The gauge change in $\delta\phi(x^i)$ is (to first order)
\begin{equation}
\delta[\delta\phi(x^i)] = \phi_{0}'(y)\xi^{5}(x^i)
\end{equation}
We see that $h_{ij}^{TT}$ (which contains the Kaluza-Klein modes) is gauge invariant, and Eq.(20) implies that $f^T$ and $\delta\phi$ are invariant on the branes at $y_1$ and $y_2$.  Eq.(22) shows explicitly that it is not possible to choose a gauge function $\xi^5$ obeying Eq.(20) that sets $h_{55}$ to zero everywhere since integrating Eq.(22) to try to do this one has only one constant of integration to satisfy the two boundary conditions of Eq.(20).  As discussed in \cite{ad}, it is possible however to choose a $\xi^5$ that sets $h_{55}$ to zero on each brane
\begin{equation}
h_{55}(x^i,y_{\alpha}) = 0\,\,\,;\,\,\,\alpha=1,2
\end{equation}
and we will use this gauge in some of the discussions below.  

We conclude this section by recording the field equations and boundary conditions.  The 5D Einstein equations read
\begin{eqnarray}
R_{ij}^{TT}:&&(\frac{1}{2}\partial_{5}^{2} -2A'\partial_{5} + \frac{1}{2}e^{2A}\Box^2)h_{ij}^{TT} = \\\nonumber &&-\frac{e^{2A}}{M_{5}^3}\sum_{\alpha}T_{ij}^{TT}(y_{\alpha})\delta(y-y_{\alpha})\\
R_{j5}:&&\frac{1}{2}\partial_5\eta^{kl}(\partial_{j}h_{kl} - \partial_{l}h_{jk}) + \frac{3}{2}A'\partial_{j}h_{55} = -\frac{1}{M_{5}^{3}}(\partial_j\delta\phi)\phi_{o}'\\
\eta^{ij}R_{ij}:&&(\frac{1}{2}\partial_{5}^{2} - 4A'\partial_5)(\Box^2 h^L + f^T) + e^{2A}\Box^2 f^T + 2A'\partial_5 h_{55} + \\\nonumber &&\frac{e^{2A}}{2}\Box^2 h_{55} + 4h_{55}(A'' - 4A'^2) = -\frac{8}{3M_{5}^3}V'(\phi_o)\delta\phi + \\\nonumber &&\frac{1}{M_{5}^3}\sum_{\alpha}\delta(y-y_{\alpha})(\frac{Te^{2A}}{3} -\frac{4}{3}V_{\alpha}'(\phi_o)\delta\phi + \frac{2}{3}h_{55}V_{\alpha}(\phi_o))\\
R_{55}:&&(\frac{1}{2}\partial_{5}^{2} - A'\partial_5)(\Box^2 h^L + f^T) + \frac{e^{2A}}{2}\Box^2 h_{55} + 2A'\partial_5 h_{55} + \\\nonumber &&h_{55}(4A'' - 4A'^2 - \frac{(\phi_{o}')^2}{M_{5}^3}) = -\frac{2}{3M_{5}^3}V'(\phi_o)\delta\phi -\frac{2\phi_{o}'\delta\phi'}{M_{5}^3} + \\\nonumber &&\frac{1}{M_{5}^3}\sum_{\alpha}\delta(y-y_{\alpha})(\frac{Te^{2A}}{3} -\frac{4}{3}V_{\alpha}'(\phi_o)\delta\phi + \frac{2}{3}h_{55}V_{\alpha}(\phi_o))\\
\partial^i\partial^jR_{ij}:&&(\frac{1}{2}\partial_{5}^2 - \frac{5}{2}A'\partial_5)(\Box^2 h^L) + \frac{1}{2}e^{2A}\Box^2 f^T -\\\nonumber &&\frac{1}{2}A'\partial_5 f^T +\frac{1}{2}A'\partial_5 h_{55} + \frac{1}{2}e^{2A}\Box^2 h_{55} + h_{55}(A''-4A'^2) \\\nonumber && = -\frac{2}{3M_{5}^3}V'e^{-2A}\delta\phi + \\\nonumber &&\frac{1}{3M_{5}^3}\sum_{\alpha}\delta(y-y_{\alpha})(Te^{2A} -V_{\alpha}'(\phi_o)\delta\phi + \frac{1}{2}h_{55}V_{\alpha}(\phi_o))
\end{eqnarray}
In the above $T\equiv \eta^{ij}T_{ij}$.  The $\delta(y-y_{\alpha})$ terms on the right hand side of Eqs.(33) and (34) imply that the bulk solutions obey the boundary conditions
\begin{eqnarray}
(-1)^{\alpha+1}[\partial_5(\Box^2 h^L + f^T) + 8A'h_{55}]\bigg|_{y=y_{\alpha}}  \\\nonumber =\frac{1}{3M_{5}^3}[Te^{2A} - 4V_{\alpha}'\delta\phi + 2h_{55}V_{\alpha}]\bigg|_{y=y_{\alpha}}
\end{eqnarray}
or equivalently using the vacuum equations and Eq.(40) below
\begin{eqnarray}
(-1)^{\alpha+1}[\partial_5(\Box^2 h^L -\frac{1}{3} f^T)]\bigg|_{y=y_{\alpha}} = \frac{e^{2A}}{3M_{5}^3}T(y)\bigg|_{y=y_{\alpha}}
\end{eqnarray}
Eqs. (30-34) represent a complete set of Einstein equations.

The $\delta\phi$ equation reads
\begin{eqnarray}\nonumber
&&e^{2A}\Box^2\delta\phi -4A'\delta\phi' + \delta\phi'' -  V''(\phi_o)\delta\phi + h_{55}V'(\phi_o)+ \frac{1}{2}\partial_5(\Box^2 h^L + f^T)\phi_{o}' \\&&-\frac{1}{2}\phi_{o}'\partial_5 h_{55} = \sum_{\alpha}\delta(y-y_{\alpha})[\frac{1}{2}h_{55}V_{\alpha}' + V_{\alpha}''\delta\phi]
\end{eqnarray}
with boundary conditions
\begin{eqnarray}
\delta\phi'(y_{\alpha}) = (-1)^{\alpha+1}\frac{1}{2}[\frac{1}{2}V_{\alpha}'h_{55} + V_{\alpha}''\delta\phi]\bigg|_{y=y_{\alpha}}
\end{eqnarray}

\section{$R_{j5}$ Equation}

We consider first the $R_{j5}$ equation.  Inserting in Eq.(6) and the orthogonal decomposition of $h_i$, Eq.(31) becomes
\begin{equation}
\frac{1}{2}\partial_5(\partial_j f^T - \Box^2h_{j}^T) + \frac{3}{2}A'\partial_j h_{55} = -\frac{\phi_{0}'}{M_{5}^3}\partial_j\delta\phi
\end{equation}
which can be decomposed into its transverse and longitudinal parts
\begin{eqnarray}
\partial_5 f^T + 3A'h_{55} + 2\frac{\phi_{0}'}{M_{5}^3}\delta\phi = 0\\
\partial_5 h^{T}_j = 0
\end{eqnarray}
Eq.(41) implies that $h_{j}^T$ = $h_{j}^T(x^i)$ is independent of y and one may use the remaining gauge freedom of Eqs.(24) and (26) to set $h_{j}^T$ to zero,
\begin{equation}
h_{j}^T(x^i) = 0
\end{equation}
Each of the terms in Eq.(40) are gauge variant, and it is interesting to see how the gauge invariance of the sum arises.  Thus using Eqs.(22),(23), and (28), the gauge change of the left hand side (lhs) of Eq.(40) is
\begin{equation}
\delta(lhs) = \partial_y(-6A'\xi^5) + 6A'\xi^{5}_{,5} + 2\frac{(\phi_{0}')^{2}}{M_{5}^3}\xi^5
\end{equation}
or
\begin{equation}
\delta(lhs) = (-6A''+ 2\frac{(\phi_{0}')^{2}}{M_{5}^3})\xi^5
\end{equation}
Using the vacuum metric equations, Eqs.(11,12) this reduces to
\begin{equation}
\delta(lhs) = -\frac{2}{M_{5}^3}\sum_{\alpha}\delta(y-y_{\alpha})V_{\alpha}(\phi_0)\xi^5(x^i,y)
\end{equation}
which vanishes as a consequence of Eq.(20).  Thus the gauge invariance of Eq.(40) is directly related to the condition that there be no brane bending.  

Eq.(40) allows us to eliminate $h_{55}$ in terms of $f^T$ and $\delta\phi$
\begin{equation}
h_{55} = -\frac{1}{3A'}\partial_5 f^T -\frac{2}{3M_{5}^3}\frac{\phi_{0}'}{A'}\delta\phi
\end{equation}
a relation that holds thoughout the bulk.  As mentioned in Sec.2, it is possible to choose a special gauge so that $h_{55}$ vanishes on the branes, Eq.(29).  In that case one has
\begin{equation}
\partial_5 f^T(x^i,y)\bigg|_{y=y_{\alpha}} = -\frac{2}{M_{5}^3}(\phi_{0}'(y)\delta\phi(x^i,y))\bigg|_{y=y_{\alpha}}
\end{equation}
and one can eliminate $\partial_5 f^T$ in terms of $\delta\phi$ on the branes.  [An alternate possibility is to choose a special gauge such that on the branes
\begin{equation}
h_{55}\bigg|_{y=y_{\alpha}} = -\frac{2}{3M_{5}^3}(\frac{\phi_{0}'}{A'}\delta\phi)\bigg|_{y=y_{\alpha}}
\end{equation}
and then $\partial_5 f^T$ would vanish on the branes.  However, in the following we will make use of Eq.(29)].

\section{$R_{55}$ and $\eta^{ij}R_{ij}$ Equations}

Eqs.(32) and (33) give relations between $f^T$, $h^L$, and $h_{55}$.  A convenient way to analyse these is to first consider the difference Eq.(33)-Eq.(32).  Then the $\delta(y-y_{\alpha})$ terms cancel and the resulting equation
\begin{eqnarray}\nonumber
3A'\partial_5(\Box^2 h^L + f^T)-e^{2A}\Box^2 f^T + h_{55}(12A'^2 - \frac{(\phi_{0}')^{2}}{M_{5}^3}) = \\\frac{2}{M_{5}^3}V'(\phi_0)\delta\phi -\frac{2\phi_{0}'}{M_{5}^3}\delta\phi'
\end{eqnarray}
is valid both in the bulk and on the branes.  Eliminating $h_{55}$ by Eq.(46) one has
\begin{eqnarray}
\partial_5(\Box^2 h^L -\frac{1}{3} f^T) &=& \frac{1}{3A'}e^{2A}\Box^2 f^T-\frac{(\phi_{0}')^{2}}{9A'^2M_{5}^3}\partial_5 f^T \\\nonumber&&+ \frac{2}{9A'M_{5}^3}\frac{\phi_0'}{A'}(12A'^2 - \frac{(\phi_{0}')^{2}}{M_{5}^3})\delta\phi \\\nonumber&&+ \frac{2}{3A'^2M_{5}^3}V'(\phi_0)\delta\phi -\frac{2\phi_{0}'}{3M_{5}^3A'}\delta\phi'
\end{eqnarray}
One can integrate Eq.(50) to obtain $h^L$ in terms of $f^T$ and $\delta\phi$.  The boundary conditions Eq.(36) involve precisely the same combination as the l.h.s. of Eq.(50), and since Eq.(50) has been seen to hold on the branes (with no $\delta(y-y_{\alpha})$ singular terms), its solution can be inserted into Eq.(36).  In the static approximation one has \cite{ad}
\begin{eqnarray}
T(y_{\alpha}) = -T_{00}(y_{\alpha})=-e^{2A(y_{\alpha})}\bar{m}_{\alpha}\delta^3(r-r(t))
\end{eqnarray}
so that on the boundaries one has
\begin{eqnarray}\nonumber
&&[\frac{1}{3A'}e^{2A}\Box^2 f^T -\frac{(\phi_{0}')^{2}}{9A'^2M_{5}^3}\partial_5 f^T + \frac{2}{3A'M_{5}^3}(V' + \phi_{0}'(4A' - \frac{(\phi_{0}')^{2}}{3A'M_{5}^3}))\delta\phi - \\ &&\frac{2\phi_{0}'}{3A'M_{5}^3}\delta\phi']\bigg|_{y=y_{\alpha}} = \frac{(-1)^{\alpha}}{3M_{5}^3}e^{2A(y_{\alpha})}T_{00}(y_{\alpha})
\end{eqnarray}
Finally in the gauge choice of Eq.(47) this reduces to
\begin{eqnarray}
\Box^2f^T\bigg|_{y=y_{\alpha}} =&& \frac{(-1)^{\alpha}}{M_{5}^3}A'T_{00}(y_{\alpha})-\\\nonumber &&\frac{2}{M_{5}^3}[e^{-2A}\delta\phi(V'(\phi_0) + 4A'\phi_{0}' -\frac{\gamma_{\alpha}}{2}\phi_{0}')]\bigg|_{y=y_{\alpha}}
\end{eqnarray}
Note that $f^T$ and $\delta\phi$ are gauge invariant on the branes so that Eq.(53) is a gauge invariant relation.  

The quantity that governs the Newtonian potential is $h_{00}$, and in the static limit this is given on the branes by
\begin{eqnarray}
h_{00}(x^i,y_{\alpha}) = h_{00}^{TT}(x^i,y_{\alpha})-\frac{1}{3}f^T(x^i,y_{\alpha})
\end{eqnarray}
Eq.(53) determines the $f^T$ contribution in terms of $T_{00}$ and $\delta\phi$.  The effect of the scalar stabilizing term is to add an additional term, the bracket of Eq.(53), and modify the $A'$ factor in the first term, e.g. for the example of Eqs.(14) and (15)
\begin{eqnarray}
A' = \beta -\frac{b}{6}\frac{\phi_{0}^2}{M_{5}^3}
\end{eqnarray}
To examine the effects of these modifications to $f^T$ we consider next the $\delta\phi$ field equation.

\section{$\delta\phi$ Field Equation}

The $\delta\phi$ field equation, Eq.(37), depends on both $h_{55}$ and the combination $\partial_5(\Box^2 h^L + f^T)$.  One may eliminate $h_{55}$ using Eq.(46) and $\partial_5(\Box^2 h^L + f^T)$ by Eq.(50).  One gets in this way a rather complicated equation involving only $\delta\phi$ and $f^T$.  While $f^T$ is determined on the branes by Eq.(53) (and is gauge invariant there), it is gauge variant in the bulk, as is $\delta\phi$.  However from Eqs.(23) and (28), the combination
\begin{eqnarray}
Q \equiv \delta\phi + \frac{1}{6}\frac{\phi_{0}'}{A'}f^T
\end{eqnarray} 
is gauge invariant in the bulk.  Thus if we eliminate $\delta\phi$ in terms of $Q$, one will obtain an equation involving only $Q$, $f^T$, $\partial_5 f^T$, and $\partial_{5}^{2}f^T$.  However, the latter three are gauge variant in the bulk, and so gauge invariance implies that the coefficients of these three quantities must actually vanish leaving an equation involving only the gauge invariant quantity $Q$.  A detailed and somewhat lengthy calculation shows that this is indeed the case and the equation for $Q$ reduces to the following relatively simple form in the bulk
\begin{eqnarray}\nonumber
e^{2A}\Box^2Q + Q'' -4A'Q' -V''Q + \frac{4}{3M_{5}^3}(2(\phi_{0}')^2 + \frac{\phi_{0}'}{A'}V')Q -\frac{2(\phi_{0}')^4}{9A'^{2}M_{5}^6}Q\\ = 0
\end{eqnarray}
Eq.(57) thus gives us an uncoupled equation that determines $Q$ in the bulk.  One may limit the solution to the branes and impose the boundary conditions of Eq.(38).  In terms of Q, this reads
\begin{eqnarray}
&&[Q' -\frac{1}{6}(\frac{\phi_{0}'}{A'})'f^T-\frac{1}{6}\frac{\phi_{0}'}{A'}\partial_5 f^T]\bigg|_{y=y_{\alpha}} =\\\nonumber && \frac{1}{2}(-1)^{\alpha+1}[\frac{1}{2}V_{\alpha}'h_{55} + V_{\alpha}''Q - \frac{1}{6}V_{\alpha}''(\frac{\phi_{0}'}{A'})f^T]\bigg|_{y=y_{\alpha}}
\end{eqnarray}
One needs
\begin{equation}
(\frac{\phi_{0}'}{A'})' = \frac{\phi_{0}''}{A'} - \frac{\phi_{0}'}{A'^2}A''
\end{equation}
and using Eqs.(11-13)
\begin{eqnarray}
(\frac{\phi_{0}'}{A'})' = (4\phi_{0}' + \frac{V'}{A'} - \frac{1}{3M_{5}^3}\frac{(\phi_{0}')^3}{A'^2}) + \sum_{\alpha}\delta(y-y_{\alpha})(\frac{V_{\alpha}'}{A'}-\frac{1}{3M_{5}^3}\frac{(\phi_{0}')V_{\alpha}}{A'^2})
\end{eqnarray}
We interpret the prescription of Eq.(58) to mean the limit from the bulk as $y \rightarrow y_{\alpha}$ and so the last term in Eq.(60) does not contribute to Eq.(58).  One has then the boundary condition
\begin{eqnarray}
&&[Q' -\frac{1}{6}(4\phi_{0}' + \frac{V'}{A'}- \frac{1}{3M_{5}^3}\frac{(\phi_{0}')^3}{A'^2})f^T - \frac{1}{6}\frac{\phi_{0}'}{A'}\partial_5 f^T]\bigg|_{y=y_{\alpha}} = \\\nonumber &&\frac{1}{2}(-1)^{\alpha+1}[[\frac{1}{2}V_{\alpha}'h_{55} + V_{\alpha}''Q - \frac{1}{6}\frac{\phi_{0}'}{A'}V_{\alpha}''f^T]\bigg|_{y=y_{\alpha}}
\end{eqnarray}
It is convenient now to make use of the gauge condition Eq.(29) which also allows us to eliminate $\partial_5 f^T(x^i,y_{\alpha})$ by Eq.(47) (from the $R_{j5}$ field equation)
\begin{eqnarray}
\partial_5 f^T\bigg|_{y=y_{\alpha}} = -\frac{2}{M_{5}^3}[\phi_{0}'(Q-\frac{1}{6}\frac{\phi_{0}'}{A'}f^T)]\bigg|_{y=y_{\alpha}}
\end{eqnarray}
The Q boundary condition then becomes
\begin{eqnarray}\nonumber
&&[Q' -\frac{1}{6}(4\phi_{0}' + \frac{V'}{A'}- \frac{1}{3M_{5}^3}\frac{(\phi_{0}')^3}{A'^2})f^T + \frac{1}{3M_{5}^3}\frac{(\phi_{0}')^2}{A'}(Q-\frac{1}{6}\frac{\phi_{0}'}{A'}f^T)]\bigg|_{y=y_{\alpha}} =\\&& \frac{1}{2}(-1)^{\alpha+1}[V_{\alpha}''Q - \frac{1}{6}\frac{\phi_{0}'}{A'}V_{\alpha}''f^T]\bigg|_{y=y_{\alpha}}
\end{eqnarray}
We can also eliminate $\delta\phi$ in terms of $Q$ in Eq.(53) yielding
\begin{eqnarray}\nonumber
&&\Box^2f^T\bigg|_{y=y_{\alpha}} = \frac{(-1)^{\alpha}}{M_{5}^3}A'T_{00}(y_{\alpha}) + \\&&\frac{2}{M_{5}^3}[e^{-2A}(\frac{1}{2}(-1)^{\alpha+1}\phi_{0}'V_{\alpha}''-V'-4A'\phi_{0}')(Q-\frac{1}{6}\frac{\phi_{0}'}{A'}f^T)]\bigg|_{y=y_{\alpha}}
\end{eqnarray}

\section{Leading Order Solutions}

As discussed above the Newtonian potential is obtained from the static approximation to $h_{00}(x^i,y_{\alpha})$ on the branes, given in Eq.(54).  $h_{00}^{TT}(x^i,y_{\alpha})$ is to be obtained by solving Eq.(30) (which is similar to the result of \cite{ad} when no scalar field was present except for the modification of $A(y)$).  $f^T(x^i,y_{\alpha})$ on the branes is governed by the coupled equations Eqs.(63) and (64).  Since Eq.(63) depends on $Q'$, one cannot use it to eliminate $Q$ in Eq.(64) (to obtain an equation depending only on $f^T$) and one must first solve Eq.(57) for $Q$ in the bulk and then insert it into the boundary conditions Eqs.(63) and (64) and use those to determine $f^T(x^i,y_{\alpha})$ on the branes.  (Thus it is the boundary conditions on the branes that couple $Q$ and $f^T$.)  The Newtonian potential then arises from the $1/r$ part of $h_{00}(x^i,y_{\alpha})$.  Since both Eq.(30) and Eq.(57) are decoupled equations, the above analysis is in principle
  doable.

An analytic solution of the second order differential equations Eqs.(30) and (57) is not possible due to the fact they depend on the complicated functions $A(y)$ and $\phi_{0}(y)$.  We note, however, that the corrections to $A(y)$ is proportional to $\phi_{0}^2/M_{5}^3$.  On the $y_2$ brane this contains a factor of $e^{-2\beta y_2} \approx 10^{-32}$ and is very small.  If we assume also $\phi_{1}^2/M_{5}^3 \ll 1$, this correction is also small on the $y_1$ brane and one can consider an iteration scheme based on the smallness of $\phi_{0}^2/M_{5}^3$.  Thus to lowest order, Eq.(30) gives rise to the same gravitational potential as in \cite{ad} (where no scalar field was present)
\begin{eqnarray}
V^{TT}(y_{\alpha}) = -\frac{4}{3}\frac{\beta}{8\pi M_{5}^3}\frac{1}{r}[\bar{m_{\alpha}}\bar{m_{\alpha}}' + \bar{m_1}\bar{m_2}]
\end{eqnarray}
where $\bar{m_{\alpha}} = e^{-\beta y_{\alpha}}m_{\alpha}$ are the observed masses on the $y_{\alpha}$ brane (and $m_{\alpha} \approx \mathcal{O}(M_{Pl})$).  Higher order effects are presumably small since they are scaled by $\phi_{0}^2/M_{5}^3$.

Eq.(57), which determines $Q$, shows a similar structure.  Thus in the example of Eq.(16), $V''$ begins as a constant with $\mathcal{O}(\phi_{0}^2/M_{5}^3)$ corrections and the remaining terms have $\mathcal{O}(\phi_{0}^2/M_{5}^3)$, $\mathcal{O}(\phi_{0}^4/M_{5}^6)$, ... corrections.  Thus to lowest order, in the bulk $Q$ obeys the equation
\begin{eqnarray}
e^{2\beta y}\Box^2 Q + Q'' -4\beta Q' -\gamma^2 Q = 0
\end{eqnarray}
where 
\begin{equation}
\gamma^2 = [V''(\phi_{0})]_{\phi_{0}=0}\,\,\,;\,\,\,\beta = [A']_{\phi_{0}=0}
\end{equation}
(and $\gamma^2 = b^2 + 4b\beta = \mathcal{O}(M_{Pl}^2)$ in the model of Eq.(16)).  Again, higher order corrections should be of $\mathcal{O}(\phi_{0}^2/M_{5}^3)$ and be small.  The boundary conditions Eq.(63) contain corrections to the zeroth order part of $\mathcal{O}(\phi_{0}'f^T)$ as well as $\mathcal{O}((\phi_{0}')^3/M_{5}^3)f^T)$, $\mathcal{O}(((\phi_{0}')^5/M_{5}^6)Q)$.  If one neglects all these corrections, Eq.(66) represents free waves propagating in the bulk (discussed in some detail in \cite{dewolfe}).  They in general give no $1/r$ Newtonian contribution (without extreme fine tuning of parameters).  Alternately, one might impose a Sommerfeld boundary condition that requires excitations in the bulk to arise from matter on the branes.  To first order this will occur by including the $\mathcal{O}(\phi_{0}'f^T)$ term, since to zero'th order $f^T$ is proportional to $T_{00}$.  We thus take as the first order boundary condition
\begin{eqnarray}
[Q'-\frac{1}{2}(-1)^{\alpha+1}\gamma_{\alpha}Q]\bigg|_{y_{\alpha}} = -\frac{1}{6}[{\frac{\gamma^2}{\beta b} - 4 +\frac{1}{2}(-1)^{\alpha+1}\frac{\gamma_{\alpha}}{\beta}}]b\phi_0 f^T\bigg|_{y_{\alpha}}
\end{eqnarray}
where
\begin{eqnarray}
\gamma_{\alpha} \equiv V_{\alpha}''(\phi_{0})\bigg|_{\phi_{0}=0}\,\,\,;\,\,\,b \equiv -\frac{\phi_{0}'}{\phi_{0}}
\end{eqnarray}
Turning to the second boundary equation Eq.(64) one might at first suggest that one could ignore the bracket on the r.h.s. as it is $\mathcal{O}((\phi_{0}^2/M_{5}^3)f^T)$, and include its effects in by iteration.  However, to lowest order, the $T_{00}$ term gives a Newtonian piece to $f^T \sim 1/\mathnormal{r}$, and if one inserts that into the bracket term, one sees that in the static limit ($\Box^2 \rightarrow \nabla^2$) the next approximation goes as $\nabla^{-2}(1/r)$ which is infrared divergent.  Thus one must include the lowest order part of the bracket in the first approximation:
\begin{eqnarray}
&&\Box^2f^T\bigg|_{y=y_{\alpha}} = \frac{(-1)^{\alpha}}{M_{5}^3}\beta T_{00}(y_{\alpha}) + \\\nonumber &&\frac{2}{M_{5}^3}[e^{-2\beta y}(4\beta b -\frac{1}{2}(-1)^{\alpha+1}\gamma_{\alpha}b-\gamma^2)\phi_0(Q+\frac{1}{6}\frac{b\phi_{0}}{\beta}f^T)]\bigg|_{y=y_{\alpha}}
\end{eqnarray}
We have kept the $Q$ term on the r.h.s. of Eq.(70) as we will see below it is of size $\phi_{0}f^T$ as a consequence of Eq.(68).  

\section{First Order Solutions for $Q$ and $f^T$}

To obtain the $f^T$ part of the Newtonian potential to first order one must solve Eq.(66), insert it into the coupled boundary conditions Eqs.(68) and (70), and then solve these equations for $f^T(y_{\alpha})$.  To solve Eq.(66) we let
\begin{equation}
Q(x^i,y) = \int d^4pe^{ipx}Q(p^i,y)
\end{equation}
and set 
\begin{equation}
Q(p,y) = e^{2\beta y}R(p,\xi) \,\,\,;\,\,\, \xi(y) = \frac{m}{\beta}e^{\beta y}
\end{equation}
where $m^2 = -p^2 = (p^0)^2 - \vec{p}^2$.  (In Eq.(72) $m/\beta$ is shorthand for $(m^2/\beta^2)^{1/2}$ where the branch cut is defined by being real and positive for $m^2 > 0$.) One finds that $R$ obeys the Bessel equation and so the general solution to Eq.(66) is
\begin{eqnarray}
Q(p,y) = e^{2\beta y}[A(p)J_{\nu}(\xi) + B(p)N_{\nu}(\xi)]
\end{eqnarray}
where $\nu = (4+ \gamma^2/\beta^2)^{1/2}$, and $A$ and $B$ are constants of integration to be determined by Eq.(68).

It is convenient to introduce the notation
\begin{equation}
\lambda_b = b/\beta
\end{equation}
Since by hypothesis $b$ and $\beta$ are $\mathcal{O}(M_{Pl})$ we expect $\lambda_b = \mathcal{O}(1)$.  We require $b > 0$ so that $\phi_{0}(y)$ is a decreasing function, and $\beta > 0$ to achieve the gauge hierarchy so that $\lambda_b > 0$.  With this notation $\gamma^2/\beta^2  = \lambda_{b}^2 + 4\lambda_b$ in the model of Eq.(16) so that
\begin{equation}
\nu = 2 + \lambda_b > 2
\end{equation}
(Appendix A shows that the above actually represents the leading terms of a general model when $\phi_{0}^2(y_1)/M_{5}^3$ is small and $\phi_{0}(y)$ is a decreasing function.)  Imposing the boundary conditions Eq.(68) determines $A$ and $B$ to be
\begin{eqnarray}
A &=& \frac{1}{D}[\xi_2N_{\nu-1}(\xi_2) - (\nu - 2 -\lambda_2)N_{\nu}(\xi_2)][-\frac{(\lambda_b + \lambda_1)}{6\beta}\phi_0'(y_1)f^T(y_1)]\\\nonumber&-&\frac{1}{D}[\xi_1N_{\nu-1}(\xi_1) - (\nu - 2 + \lambda_1)N_{\nu}(\xi_1)][-\frac{e^{2\beta y_2}}{6\beta}(\lambda_b - \lambda_2)\phi_0'(y_2)f^T(y_2)]
\end{eqnarray}
and
\begin{eqnarray}
B &=& -\frac{1}{D}[\xi_2J_{\nu-1}(\xi_2) - (\nu - 2 -\lambda_2)J_{\nu}(\xi_2)][-\frac{(\lambda_b + \lambda_1)}{6\beta}\phi_0'(y_1)f^T(y_1)]\\\nonumber&+&\frac{1}{D}[\xi_1J_{\nu-1}(\xi_1) - (\nu - 2 + \lambda_1)J_{\nu}(\xi_1)][-\frac{e^{2\beta y_2}}{6\beta}(\lambda_b - \lambda_2)\phi_0'(y_2)f^T(y_2)]
\end{eqnarray}
where $\xi_{1,2} = \xi(y_{1,2})$ and D is given by
\begin{eqnarray}\nonumber
D = [\xi_1J_{\nu-1}(\xi_1) - (\nu - 2 + \lambda_1)J_{\nu}(\xi_1)][\xi_2N_{\nu-1}(\xi_2) - (\nu-2-\lambda_2)N_{\nu}(\xi_2)]\\-[\xi_1N_{\nu-1}(\xi_1) - (\nu - 2 + \lambda_1)N_{\nu}(\xi_1)][\xi_2J_{\nu-1}(\xi_2) - (\nu - 2 - \lambda_2)J_{\nu}(\xi_2)]
\end{eqnarray}

The boundary condition Eq.(70) reduces to leading order in momentum space to
\begin{eqnarray}
&&m^2 f^T(y_{\alpha}) = (-1)^{\alpha}\frac{\beta}{M_{5}^3}T_{00}(y_{\alpha})  \\\nonumber &&+ 2\frac{\beta b}{M_{5}^3}[(-1)^{\alpha}\lambda_{\alpha}-\lambda_b]\phi_0[(AJ_{\nu}(\xi) + BN_{\nu}(\xi)) + \frac{\lambda_b}{6}e^{-2\beta y}\phi_0 f^T(y)]\bigg|_{y=y_{\alpha}}
\end{eqnarray}
where 
\begin{eqnarray}
\lambda_{\alpha} \equiv \frac{\gamma_{\alpha}}{2\beta} = \mathcal{O}(1)
\end{eqnarray}
To calculate the Newtonian contribution of $f^T(y_{\alpha})$ we take the static limit ($m^2 = -p^2 \rightarrow -\vec{p}^2$) and take the low momentum limit of the right hand side (rhs) by limiting $\xi_{1,2}\rightarrow0$.  For $\xi_2$ this means we assume
\begin{eqnarray}
\xi_2 = \frac{m}{\beta}e^{\beta y_2} \ll 1
\end{eqnarray}
or we are considering momenta
\begin{eqnarray}
p \leq \beta e^{-\beta y_2} \approx M_{Pl}10^{-16} \approx 1TeV
\end{eqnarray}
corresponding to distances $r \gtrsim 10^{-17}$cm.  (The expansion for $\xi_1$ is valid for distances greater than the Planck length).  Thus for all experimental tests of Newtonian forces, this expansion is valid.  Keeping the leading terms we find for $f^T(y_2)$ that
\begin{eqnarray}
&&AJ_{\nu}(\xi_2) + BN_{\nu}(\xi_2) \simeq\\\nonumber&& \frac{1}{D}[\frac{(\nu- 2 -\lambda_2)\lambda_b}{6\nu\pi}\phi_1(\lambda_b + \lambda_1)f^T(y_1)\\\nonumber&&- \frac{(\nu - 2 + \lambda_1)\lambda_b}{6\nu\pi}(\lambda_b-\lambda_2)\phi_1f^T(y_2)\\\nonumber&& + \frac{(\nu + 2 + \lambda_2)\lambda_b}{6\nu\pi}\phi_1(\lambda_b + \lambda_1)f^T(y_1)\\\nonumber&&- \frac{(\nu + 2 - \lambda_1)\lambda_b}{6\nu\pi}(\lambda_b-\lambda_2)\phi_1f^T(y_2)e^{-2\nu\beta y_2}]
\end{eqnarray}
where expanding $1/D$ gives
\begin{eqnarray}
1/D \cong -\frac{\nu\pi e^{-\nu\beta y_2}}{(\nu + 2 + \lambda_2)(\nu-2+\lambda_1)}[1 + \frac{(\nu + 2 - \lambda_1)(\nu - 2 - \lambda_2)}{(\nu - 2 + \lambda_1)(\nu + 2 + \lambda_2)}e^{-2\nu\beta y_2}]
\end{eqnarray}
Eq.(79) in the static limit then becomes to leading order (the second term of Eq.(84) is negligible)
\begin{eqnarray}
-\vec{p}^2f^{T}(y_2) = \frac{\beta}{M_{5}^3}T_{00}(y_2) + \alpha^2 e^{-(2+2\lambda_b)\beta y_2}[f^T(y_2)-f^T(y_1)]
\end{eqnarray}
where
\begin{eqnarray}
\alpha^2 = \frac{2}{3}\frac{\phi_{1}^2}{M_{5}^3}\beta^2\lambda_{b}^2\frac{(2+\lambda_b)(\lambda_2-\lambda_b)}{4+\lambda_b + \lambda_2}
\end{eqnarray}
The $e^{-(2+2\lambda_b)\beta y_2}$ in Eq.(85) comes from the $e^{-\nu\beta y_2}$ of $1/D$ and the $\phi_{0}$ factor in Eq.(70) (i.e. $\phi_0 = \phi_1 e^{-\lambda_b\beta y_2}$).

The analysis for $f^T(y_1)$ is more subtle.  Here we find that the numerator of $AJ_{\nu}(\xi_1)+BN_{\nu}(\xi_1)$ terms of $Q$ gives contributions of size $e^{-\nu\beta y_2}$ (from e.g. $N_{\nu}(\xi_2)J_{\nu}(\xi_1)\sim(\xi_1/\xi_2)^{\nu}$) and size $e^{\nu\beta y_2}$ (from $J_{\nu}(\xi_2)N_{\nu}(\xi_1)$).  Multiplying by $1/D$ then gives terms of size $e^{-2\nu\beta y_2}$ and $\mathcal{O}(1)$.  The $\mathcal{O}(1)$ term actually cancels with the $\mathcal{O}(1)$ term of $\phi_0f^T(y_1)/6$ of Eq.(79), and so one must keep the second factor in the bracket of Eq.(84) to get a total result of size $e^{-2\nu\beta y_2}$ on the rhs:
\begin{eqnarray}
-\vec{p}^2f^T(y_1) = -\frac{\beta}{M_{5}^3}T_{00}(y_1) + \alpha^2 e^{-(4+2\lambda_b)\beta y_2}(f^T(y_2) - f^T(y_1))
\end{eqnarray}
Note that $\lambda_1$ does not enter in these leading order results.

One can now easily solve Eqs.(85) and (87) to get in coordinate space the results
\begin{eqnarray}
-\frac{1}{3}f^T(y_2) = \frac{1}{3}\frac{G_N\bar{m_2}}{r}e^{-\mu r}e^{2\beta y_2} -\frac{1}{3}\frac{G_N}{r}(m_1 + \bar{m_2})(1-e^{-\mu r})
\end{eqnarray}
and
\begin{eqnarray}\nonumber
-\frac{1}{3}f^T(y_1) = -\frac{1}{3}\frac{G_N}{r}(m_1 + \bar{m_2}) + \frac{1}{3}\frac{G_N}{r}\bar{m_2}e^{-\mu r} \\- \frac{1}{3}\frac{G_N}{r}e^{-2\beta y_2}(m_1 + \bar{m_2})(1-e^{-\mu r})
\end{eqnarray}
where the Newton constant is given by
\begin{eqnarray}
G_N \equiv \frac{\beta}{8\pi M_{5}^3}
\end{eqnarray}
and\footnote{In the ``stiff potential'' limit, $\lambda_2 \rightarrow \infty$, Eq.(91) reduces to the mass of the radion given in Eq.(6.6) of \cite{csaki2} and Eq.(3.19) of \cite{tm}, and for general $\lambda_2$ it is equal to the mass of Eq.(30) of \cite{tk2}.  These papers are discussed further in Appendix B.}
\begin{eqnarray}
\mu^2 \equiv \alpha^2 e^{-(2+2\lambda_b)\beta y_2}
\end{eqnarray}
The requirement $\mu^2 \geq 0$ implies $\lambda_2 \geq \lambda_b$ or $\lambda_2 < -(4+\lambda_b)$.

In the limit $\phi_1 \rightarrow0$ (no scalar field) Eqs.(88) and (89) reduce to the results of \cite{ad}.  The presence of the scalar field does indeed grow a mass for the $f^T$ field and so in the limit $\mu r \gg 1$ the remaining $1/r$ piece in Eq.(88) precisely combines with the $h_{00}^{TT}$ of Eq.(65) to give a total Newton potential with Newton constant of Eq.(90) on the TeV brane.  (On the Planck brane the last term in Eq.(89) is negligible and one gets an additional factor of 5/3.)  However, the factor $e^{2\beta y_2}$ in the first term of Eq.(88) remains, and the mass $\mu$ is suppressed by $e^{-(1+\lambda_b)\beta y_2}$.  One may ask how large r has to get so that this anomalous behavior becomes negligible and Newtonian physics is reproduced on the TeV brane.  As a measure of the effects seen here, we assume that the Newtonian force has been measured at the 1\% level, so that the large dominant term in Eq.(88) implies
\begin{eqnarray}
\frac{1}{3}e^{-\mu r}e^{2\beta y_2} < 10^{-2}
\end{eqnarray}
or
\begin{eqnarray}
\mu r > 2\beta y_2 - ln(0.03)
\end{eqnarray}
Since we are assuming $\phi_1$ is small and $\beta \simeq M_{Pl}$ we set $\phi_{1}^2/M_{5}^3 = 1/10$, $\beta = 1.22\times10^{19} GeV$ and $e^{2\beta y_2} = 10^{32}$.  Eq.(93) then implies
\begin{eqnarray}
f^{1/2}e^{-\lambda_b\beta y_2}r > 4.827\times10^{-15}cm
\end{eqnarray}
where
\begin{eqnarray}
f = \frac{\lambda_{b}^2(2+\lambda_b)(\lambda_2-\lambda_b)}{4+\lambda_b+\lambda_2}
\end{eqnarray}
Current gravitational force experiments have been done at a separation between masses as small as $10\mu m$\cite{long}.
Hence we require
\begin{eqnarray}
f^{1/2}e^{-\lambda_b\beta y_2} > 4.827\times10^{-12}
\end{eqnarray}
Eq.(96) gives an exclusion contour in the $\lambda_2 - \lambda_b$ parameter space.  For example for $\lambda_2 = 1$, one requires
\begin{eqnarray}
\lambda_b < 0.67
\end{eqnarray}
\begin{figure}[h]
\begin{center}
\includegraphics{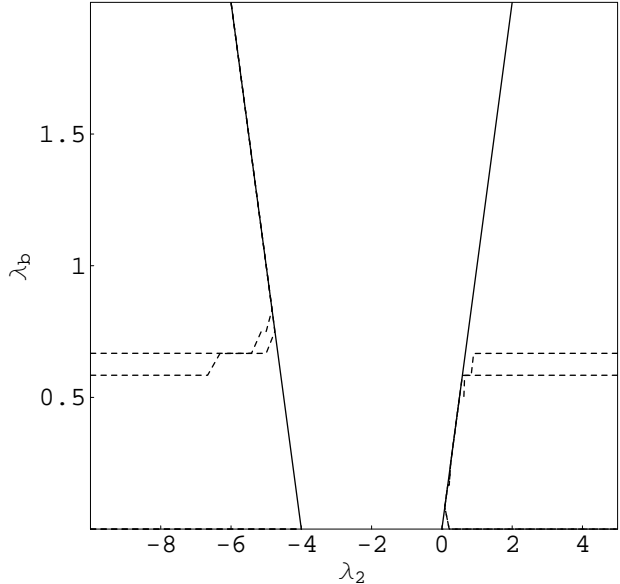}
\end{center}
\caption{Allowed region in the $\lambda_b -\lambda_2$ parameter space for $\beta =M_{Pl}$.  Allowed points must lie below the solid lines where $\mu^2$ (defined in Eqs.(91) and (86)) is correctly positive.  The further restriction arising from the measurement \cite{long} of the Newtonian force down to $10\mu m$, Eq.(96), is given by the upper dashed curves, the allowed region lying below them.  (The lower dashed curves represent the restriction if data were improved to verify the Newtonian force down to $1\mu m$.)  In general $\lambda_b$ and $\lambda_2$ (Eqs.(74)and(80)) are expected to be $\mathcal{O}(1)$.} 
\end{figure}
\begin{figure}[h]
\begin{center}
\includegraphics{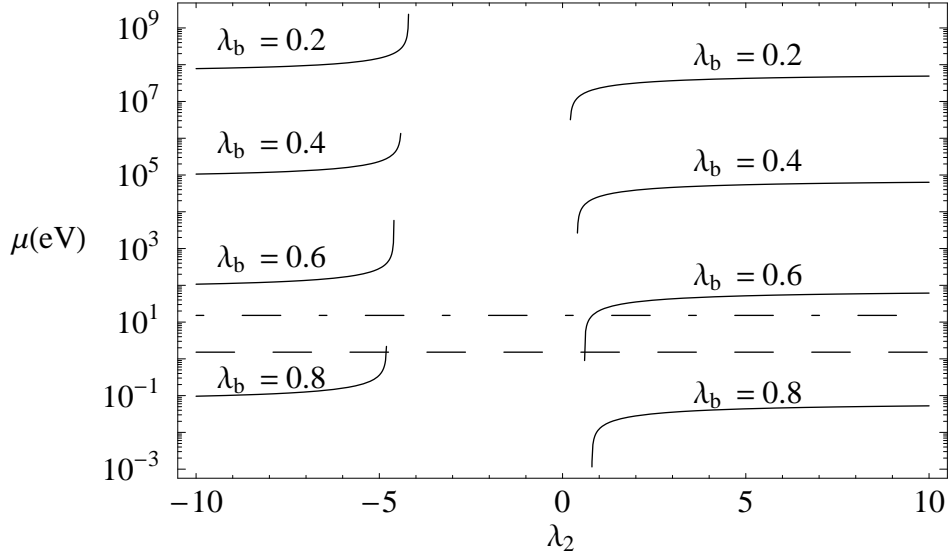}
\end{center}
\caption{$\mu$ of Eqs.(86) and (91) as a function of $\lambda_2$ and $\lambda_b$.  The lower dashed curve is the lower bound for $\mu$ for the current data $r=10\mu m$.  The upper dashed curve would be the bound if experiments were improved by a factor of 10.  The central region is excluded by the constraint $\mu^2$$\ge 0$.} 
\end{figure}
to avoid disagreement with experiment.  The general excluded region is shown in Fig.1.  Thus the absence of any deviation from the Newtonian force law already rules out a large amount of parameter space.  Fig.2 plots the mass $\mu$ as a function of $\lambda_2$ for various values of $\lambda_b$.  $1/\mu$ represents the range of the Yukawa-like potential of $f^T$ in Eq.(88) (1/eV $\cong 1.97\times 10^{-4}$mm).  However, the range by itself does not govern the strength of the interaction due to the anomalous $e^{2\beta y_2}$ factor in the first term of Eq.(88).  Fig.2 illustrates the interplay of these two factors.  Thus for $\lambda_b$ small, $\mu$ is large and the exponential decay suppresses the $e^{2\beta y_2}$ enhancement at $r \geq 10\mu m$.  However, as $\lambda_b$ grows, $\mu$ decreases until finally the $e^{2\beta y_2}$ factor causes a violation of the current data \cite{long}.  The lower dashed curve represents the lower limit on $\mu$ from Eq.(93) for the current experimental data at $r=10\mu m$.  The upper dashed curve would be the bound if experiments were improved and obtained negative results at $r=1\mu m$.  Note from Eq.(86) that $\mu^2 \rightarrow \infty$ as $\lambda_2 \rightarrow -(4+\lambda_b)$, and $\mu^2 \rightarrow 0$ as $\lambda_2 \rightarrow \lambda_b$, as can be seen in Fig. 2.


\section{Coupling of the Goldberger-Wise Field to the Standard Model}
The action of Eq.(9) couples the scalar Goldberger-Wise (GW) field $\phi$ to the gravitational field on the branes, which is necessary in order to cancel the cosmological constant.  This also leads to an indirect non-local coupling of $\phi$ to Standard Model matter fields via the coupled boundary conditions Eqs.(68) and (70).  We consider here the possibility of a direct coupling of $\phi$ to Standard Model matter fields.  To illustrate what might happen in this case, we consider the simple example of adding to the Lagrangian of Eq.(9) the interaction on the TeV brane
\begin{eqnarray}
\mathcal{L}_{\phi} = \sqrt{-^4g}\frac{\lambda}{M_{5}}\phi^2(x^i,y_2)\chi^2(x^i)
\end{eqnarray}
where $\chi$ is the scalar field of Eq.(3) (e.g. perhaps the Higgs field) with an electroweak sized observed mass $\bar{m}$.  Here $\lambda \lesssim \mathcal{O}(1)$ is a coupling constant and we have scaled the interaction by $M_5 =\mathcal{O}(M_{Pl})$ since $\phi$ is a 5D field of dimension [mass$]^{3/2}$ (and in RS models only Planck size masses enter fundamental interactions).

The effect of Eq.(98) on the gravitational fields is simply to add this interaction to the stress tensor.  To see the effects on $\phi$, we expand to quadratic order in the deviation from the vacuum metric where $h_{ij},\delta\phi$, and $\chi^2$ are treated as first order.  Then 
\begin{eqnarray}
\mathcal{L}_{\phi} = \frac{\lambda}{M_{5}}e^{-4A}[\phi_{o}^2\chi^2 + \frac{1}{2}\eta^{ij}h_{ij}\phi_{o}^2\chi^2 + 2\phi_o\delta\phi\chi^2]
\end{eqnarray}
where the omitted terms are of cubic or higher order.  We consider the situation for the leading terms of Eqs.(14-18) where $\phi_{1}^2/M_{5}^3$ is small.  Then replacing $\chi$ by its canonically normalized field $\chi' = e^{-\beta y_2}\chi$, the first term becomes an additional mass to the $\chi'$ field
\begin{eqnarray}
\mathcal{L}_{\phi}^{(1)} = \lambda\frac{\phi_{1}^2}{M_{5}^3}\tilde{m}^2e^{-2\lambda_b\beta y_2}\chi'^2
\end{eqnarray}
where
\begin{eqnarray}
\tilde{m} \equiv M_5e^{-\beta y_2} = \mathcal{O}(\mathrm{TeV})
\end{eqnarray}
We see that the additional mass is suppressed by the factor $e^{-\lambda_b\beta y_2}$ and hence is negligible.  Similarly, the second term just couples this negligible mass to the gravitational field.

Using Eq.(56) the third term of Eq.(99) becomes
\begin{eqnarray}
\mathcal{L}_{\phi}^{(3)} = 2\lambda\bigg(\frac{\phi_{1}^2}{M_{5}^3}\bigg)^{1/2}\frac{\tilde{m}^2}{M_{5}^{3/2}}e^{-\lambda_b\beta y_2}Q\chi'^2 + \frac{1}{3}\lambda\bigg(\frac{\phi_{1}^2}{M_{5}^3}\bigg)\tilde{m}^2e^{-2\lambda_b\beta y_2}\lambda_bf^T\chi'^2
\end{eqnarray}
The last term of Eq.(102) is the same size as the second term in Eq.(99).  Thus the only term in Eq.(99) that needs further discussion is the first term of Eq.(102) and so we may write
\begin{eqnarray}
\mathcal{L}_{\phi} \cong 2\lambda\bigg(\frac{\phi_{1}^2}{M_{5}^3}\bigg)^{1/2}\frac{e^{-\lambda_b\beta y_2}}{M_{5}^{3/2}}Q(x^i,y_2)[\tilde{m}^2\chi'^2]
\end{eqnarray}
Eq.(103), which is a 4D Lagrangian on the $y_2$ brane implies an additional term proportional to a 'mass' stress tensor contribution $\tilde{m}^2\chi'^2$ to be added to the right hand side of Eq.(68)
\begin{eqnarray}
\beta\phi_o\bigg(\frac{M_5}{\beta}\bigg)2\lambda\frac{\tilde{m}^2\chi'^2}{M_{5}^4}\bigg|_{y_2}
\end{eqnarray}
In the static limit this term is clearly suppressed compared to the $f^T$ term in Eq.(68) since $\tilde{m}^2\chi'^2 \thickapprox \mathcal{O}[(TeV)^4]$ and the $f^T$ factor has the additional $e^{2\beta y_2}$ enhancement.

It remains therefore to discuss the dynamical effects implied by Eq.(103) arising from the Kaluza-Klein (KK) modes of $Q$.  These are a set of discrete modes with $m_n \equiv (-p_{n}^2)^{1/2}$ obeying the homogeneous equations of Eqs.(66), (68), and (70) with no brane matter present (i.e. $T_{\mu\nu} = 0$).  They presumably form a complete set so that in the bulk one can write the general solution of Eq.(66) using Eq.(73) as 
\begin{eqnarray}
Q = e^{2\beta y}\sum_n[A_n(p)J_{\nu}(\xi_{n}) + B_n(p)N_{\nu}(\xi_{n})]
\end{eqnarray}
where 
\begin{eqnarray}
\xi_n(y) = \frac{m_n}{\beta}e^{\beta y}
\end{eqnarray}
The KK masses $m_n$ and the ratio $B_n/A_n$ are determined by imposing the brane boundary conditions.  We consider first the $f^T$ boundary conditions.  For the low $m_n$ modes we may use Eqs.(85) and (87) with $-p^2 \rightarrow m_{n}^2$ and $T_{00}$ set to zero:
\begin{eqnarray}
m_{n}^2 f^T(y_2) &=& \mu^2[f^T(y_2) -f^T(y_1)]\\
m_{n}^2 f^T(y_1) &=& e^{-2\beta y_2}\mu^2[f^T(y_2) -f^T(y_1)]
\end{eqnarray}
Hence $f^T(y_1) = \Delta^2f^T(y_2)$ and
\begin{eqnarray}
(m_{n}^2 -\mu^2 +\Delta^2)f^T(y_2) = 0 \,\,\,;\,\,\,\Delta^2 = \frac{\mu^4e^{-2\beta y_2}}{m_{n}^2 + \mu^2e^{-2\beta y_2}}
\end{eqnarray}
Eq.(109) has two solutions.  Either
\begin{eqnarray}
m_{n}^2 = \mu^2 -\Delta^2\,\,\,\textrm{and}\,\,\, f^T(y_2) = \textrm{arbitrary}
\end{eqnarray}
or
\begin{eqnarray}
m_{n}^2 \neq  \mu^2 -\Delta^2\,\,\,\textrm{and}\,\,\, f^T(y_2) = 0
\end{eqnarray}
Since $\Delta^2$ is very small, to leading order Eq.(110) gives
\begin{eqnarray}
m_{n}^2 \cong \mu^2 \,\,\,;\,\,\,f^T(y_1) = e^{-2\beta y_2}f^T(y_2)\,\,\,;\,\,\, f^T(y_2) = \textrm{arbitrary}
\end{eqnarray}
The possibility in Eq.(111) implies from Eq.(68)
\begin{eqnarray}
[Q' -\frac{1}{2}(-1)^{\alpha + 1}\gamma_{\alpha}Q]\bigg|_{y_{\alpha}} = 0
\end{eqnarray}
The $m_n$ in this case arise from the vanishing of $D$ in Eq.(78) i.e. from the algebraic equation
\begin{eqnarray}
\alpha\delta - \gamma\beta = 0
\end{eqnarray}
and $B_n/A_n = -\alpha/\beta$, where
\begin{eqnarray}\nonumber
\alpha &=& \xi_{n1}J_{\nu-1}(\xi_{n1}) - (\nu - 2 + \lambda_1)J_{\nu}(\xi_{n1})\\\nonumber
\beta &=& \xi_{n1}N_{\nu-1}(\xi_{n1}) - (\nu - 2 + \lambda_1)N_{\nu}(\xi_{n1})\\\nonumber
\gamma &=& \xi_{n2}J_{\nu-1}(\xi_{n2}) - (\nu - 2 - \lambda_2)J_{\nu}(\xi_{n2})\\
\delta &=& \xi_{n2}N_{\nu-1}(\xi_{n2}) - (\nu - 2 - \lambda_2)N_{\nu}(\xi_{n2})
\end{eqnarray}
and
\begin{eqnarray}
\xi_{n\alpha} = \xi_{n}(y_{\alpha})\,\,\,;\,\,\,\alpha = 1,2
\end{eqnarray}
Eq.(112) corresponds to what is commonly called the "radion" mode.  However, note that this mode arises in $Q$ which is the gauge invariant part of $\delta\phi$ and is not an aspect of the metric components $f^T(x^i,y)$ or $h_{55}(x^i,y)$ which are pure gauge variant in the bulk, as seen in Eqs.(22) and (23).

In the following we will label the "radion" mode of Eq.(112) by $n = 0$, and the usual KK modes arising from Eq.(114) by $n = 1, 2, 3, \dots$.  We begin by examining the KK mode contributions to Eq.(103).

A numerical analysis is needed to get accurate solutions of Eq.(114) for $m_n$.  However, since the lowest $m_n$ is $\mathcal{O}$(TeV), one has that $\xi_{n1} \ll 1$ (provided $m_n \ll M_{Pl}$) while $\xi_{n2} > 1$, and $\xi_{n2}$ becomes large rapidly.  Thus approximate solutions of Eq.(114) can be obtained using the asymptotic forms for the Bessel functions:
\begin{eqnarray}
\mathrm{tan}(\xi_{n2} - \frac{\nu\pi}{2} + \frac{\pi}{4}) = \frac{2}{(4\nu^2 - 1) + 2(\nu - 2 - \lambda_2)}\xi_{n2}
\end{eqnarray}
One can qualitatively see where the $m_n$ lie.  Aside from exceptional values of $\nu \equiv 2 + \lambda_b$ and $\lambda_2$ (to be discussed below) the coefficient of $\xi_{n2}$ on the right is small, and so $\eta_{n}\equiv \xi_{n2} -\nu\pi/2 + \pi/4$ is close to $\pi, 2\pi, 3\pi,\ldots$.  As $n$ gets large $\xi_n$ gets large and so the postitions of $\eta_n$ migrates to be close to $(2n + 1)\pi/2$.  The exceptional situation occurs when one fine tunes $\lambda_2$ and $\lambda_b$ so that the denominator on the right nearly vanishes, making the postions of $\eta_n$ start near $\pi/2, 3\pi/2$ for small $n$ as well.

In the following analysis, we will assume
\begin{eqnarray}
\xi_{n2} \gg 1
\end{eqnarray}
and calculate the leading terms in the expansion in powers of $1/\xi_{n2}$.  This should be accurate for $n \gtrsim 3$, and give a qualitative picture for lower $n$.  We also have $\xi_{n1} \ll 1$ so that since $B_n/A_n = -\alpha/\beta$ we can write
\begin{eqnarray}
Q = \sum_n A_n(x^i)F_n(y)
\end{eqnarray} 
where ($\xi_n(y) = m_ne^{\beta y}/\beta$)
\begin{eqnarray}
F_n(y) = e^{2\beta y}[J_{\nu}(\xi_n) - \bigg(\frac{\xi_{n1}}{2}\bigg)^{2\nu}\frac{\pi}{\nu\Gamma^2(\nu)}\bigg(\frac{\nu + 2 - \lambda_1}{\nu - 2 + \lambda_1}\bigg)N_{\nu}(\xi_n)]
\end{eqnarray}
The second term in Eq.(120) is generally negligible for $m_n \ll M_{Pl}$ unless $\xi_n(y)$ is close to $\xi_{n1}$, where it becomes comparable to the first term.

We turn next to obtain the canonical normalization of the 4D fields $A_n(x^i)$.  The Lagrangian generating Eq.(66) is
\begin{eqnarray}
\mathcal{L}_{Q} = \int d^5x\sqrt{-^5g_{(o)}}[-\frac{1}{2}\partial_iQg^{ij}_{(o)}\partial_jQ - \frac{1}{2}(\partial_5Q)^2 -\frac{1}{2}\gamma^2Q^2]
\end{eqnarray}
where $g_{ij(o)}$ is the vacuum metric.  To leading order then
\begin{eqnarray}
\mathcal{L}_{Q} = -\frac{1}{2}\int d^4x\int_{y_1}^{y_2} dye^{-4\beta y}[e^{2\beta y}\partial_iQ\eta^{ij}\partial_jQ + (\partial_5Q)^2 +\gamma^2Q^2]
\end{eqnarray}
so that the kinetic energy is
\begin{eqnarray}
\mathcal{L}_{KE} = -\frac{1}{2}\int d^4x\sum_{n,m}\partial_iA_n\partial^{i}A_m\int_{y_1}^{y_2}e^{-2\beta y}F_n(y)F_m(y)
\end{eqnarray}
In general, there appears to be no orthogonality relation involving quadratic integrals of the {$F_n$}.  However, one may show that the terms with $n \neq m$ are $1/\xi_{n2}$ smaller than those with $n = m$, so that to leading order
\begin{eqnarray}
\mathcal{L}_{KE} \cong -\frac{1}{2}\sum_n\int d^4x\partial_iq_n(x)\partial^iq_n(x)
\end{eqnarray}
where
\begin{eqnarray}
A_n = \frac{q_n}{\mu_n}\,\,\,;\,\,\,\mu_{n}^2 = \int_{y_1}^{y_2}dye^{-2\beta y}[F_n(y)]^2
\end{eqnarray}
and $q_n$ are the canonically normalized fields.  In general, one must calculate the normalization factor numerically (for fixed values of $\lambda_b$ and $\lambda_1$).  However, it is possible to get an approximate analytic evaluation by changing variables from $y$ to $\xi_n$ and decomposing the integral into a low $\xi_{n}$ part and a high $\xi_n$ part
\begin{eqnarray}
\mu_{n}^2 = \frac{1}{\beta}\int_{\xi_{n1}}^{1}\frac{d\xi_n}{\xi_n}e^{-2\beta y}F_{n}^2 + \frac{1}{\beta}\int_{1}^{\xi_{n2}}\frac{d\xi_n}{\xi_n}e^{-2\beta y}F_{n}^2
\end{eqnarray}
We can then estimate the second integral by using the large $\xi_n$ asymptotic form for the Bessel functions and the first integral by using the small $\xi_n$ form.  One finds that the first integral is a factor of $1/\xi_{n2}$ smaller than the second so we may write
\begin{eqnarray}
\mu_{n}^2 \simeq \frac{2\beta}{\pi m_{n}^2}\int_{1}^{\xi_{n2}}d\xi_n\mathrm{cos}^2(\xi_n - \frac{\nu\pi}{2} - \frac{\pi}{4})
\end{eqnarray}
or
\begin{eqnarray}
\mu_{n}^2 \simeq \frac{\beta}{\pi m_{n}^2}\int_{1}^{\xi_{n2}}d\xi_n[1 + \mathrm{sin}(2\xi_n - \lambda_b\pi)]
\end{eqnarray}
Hence
\begin{eqnarray}
\mu_{n}^2 \simeq \frac{\beta}{\pi m_{n}^2}\xi_{n2}[1 - \frac{1}{2\xi_{n2}}(2 + \mathrm{cos}(2\xi_n - \lambda_b\pi) - \mathrm{cos}(2-\lambda_b\pi))]
\end{eqnarray}
Thus to leading order in $1/\xi_{n2}$ one has
\begin{eqnarray}
\mu_{n}^2 \simeq \frac{e^{\beta y_2}}{\pi m_{n}}
\end{eqnarray}
We consider next the mass terms of Eq.(121)
\begin{eqnarray}
\mathcal{L}_{\gamma}^{m} = -\frac{1}{2}\gamma^2\int d^5x\sqrt{-^5g_{(o)}}Q^2
\end{eqnarray}
and
\begin{eqnarray}
\mathcal{L}_{5}^{m} = -\frac{1}{2}\int d^5x\sqrt{-^5g_{(o)}}(\partial_5Q)^2
\end{eqnarray}
Inserting Eq.(119), and neglecting $n \neq m$ terms gives
\begin{eqnarray}
\mathcal{L}_{\gamma}^{m} = -\frac{1}{2}\gamma^2\int d^4x\int_{y_1}^{y_2}dye^{-4\beta y}\sum_n\frac{1}{\mu_{n}^2}q_{n}^2F_{n}^2
\end{eqnarray}
and hence
\begin{eqnarray}
\mathcal{L}_{\gamma}^{m} = -\frac{1}{2}\gamma^2\int d^4x\sum_nq_{n}^2\int_{\xi_{n1}^2}^{\xi_{n2}}\frac{d\xi_n}{\beta\xi_n}\frac{1}{\mu_{n}^2}e^{-4\beta y}F_{n}^2
\end{eqnarray}
Neglecting the low $\xi_n$ part gives
\begin{eqnarray}
\mathcal{L}_{\gamma}^{m} \cong -\frac{1}{2}\gamma^2\int d^4x\sum_n\frac{q_{n}^2}{\mu_{n}^2}\int_{1}^{\xi_{n2}}\frac{d\xi_n}{\beta\xi_n}J_{\nu}^2(\xi_n)
\end{eqnarray}
which evaluates to
\begin{eqnarray}
\mathcal{L}_{\gamma}^{m} \cong -\frac{1}{2}\frac{\gamma^2}{\beta}\sum_n\int d^4xe^{-\beta y_2}m_n q_{n}^2
\end{eqnarray}
Using Eqs.(67) and (16) this becomes
\begin{eqnarray}
\mathcal{L}_{\gamma}^{m} \cong -\frac{1}{2}(\lambda_{b}^2 + 4\lambda_b)\int d^4x\sum_n\frac{m_{n}^2q_{n}^2}{\xi_{n2}}[1 + \mathcal{O}(1/\xi_{n2})]
\end{eqnarray}
For Eq.(132) we use
\begin{eqnarray}
\partial_5 J_{\nu} = 2\beta J_{\nu} + \frac{1}{2}\beta\xi(J_{\nu - 1} - J_{\nu + 1})
\end{eqnarray}
Because of the extra $\xi$ factor in the second term of Eq.(138), we keep the $1/\xi_n$ correction to the leading term in the asymptotic expansion of $J_{\nu\pm 1}$ for large $\xi_n$.  Then
{\setlength\arraycolsep{.1pt}
\begin{eqnarray}\nonumber
\mathcal{L}_{5}^m \cong -\frac{1}{2}\int d^4x \sum_n \frac{q_{n}^2}{\mu_{n}^2}\int_{1}^{\xi_{n2}}d\xi_n \frac{8\beta}{\pi\xi_{n}^2}[&&(1-\frac{\nu^2}{2})\mathrm{cos}(\xi_n - \frac{\nu\pi}{2} -\frac{\pi}{4})\\ &&-\, \frac{\xi_n}{2}\mathrm{sin}(\xi_n - \frac{\nu\pi}{2} - \frac{\pi}{4})]^2
\end{eqnarray}} 
which reduces to
\begin{eqnarray}
\mathcal{L}_{5}^m \cong -\frac{1}{2}\int d^4x \sum_n \frac{q_{n}^2}{\mu_{n}^2}\frac{\beta}{\pi}\xi_{n2}[1 + \mathcal{O}(1/\xi_{n2})]
\end{eqnarray} 
Thus inserting in Eq.(130) gives
\begin{eqnarray}
\mathcal{L}_{5}^m \cong -\frac{1}{2}\int d^4x \sum_n m_{n}^2 q_{n}^2(x)[1 + \mathcal{O}(1/\xi_{n2})]
\end{eqnarray} 
Thus for large $\xi_{n2}$ the total mass term for the $q_n$ field is
\begin{eqnarray}
\mathcal{L}^m = \mathcal{L}_{5}^m + \mathcal{L}_{\gamma}^m = -\frac{1}{2}\int d^4x \sum_n m_{n}^2 q_{n}^2(x)[1 + \mathcal{O}(1/\xi_{n2})]
\end{eqnarray}
and the KK modes of $Q$ have TeV or higher masses.

We return now finally to Eq.(103) to examine the coupling of the KK modes to the SM $\chi'$ field.  Inserting Eqs.(119) and (120) gives to leading order
\begin{eqnarray}
\mathcal{L}_{\phi} \cong 2\lambda\bigg(\frac{\phi_{1}^2}{M_{5}^3}\bigg)^{1/2}e^{-\lambda_b\beta y_2}\frac{\tilde{m}^2}{M^{3/2}}e^{2\beta y_2}\sum_n\frac{q_n}{\mu_n}J_{\nu}(\xi_{n2})\chi'^2
\end{eqnarray}
where $\tilde{m}$ is defined in Eq.(101).  The expansion of the Bessel function gives
\begin{eqnarray}
J_{\nu}(\xi_{n2}) \cong \sqrt{\frac{2}{\pi\xi_{n2}}}\mathrm{sin}(\xi_{n2} - \frac{\nu\pi}{2} + \frac{\pi}{4})
\end{eqnarray}
and using Eq.(117) this becomes
\begin{eqnarray}
J_{\nu}(\xi_{n2}) \cong \sqrt{\frac{2}{\pi\xi_{n2}}}[1 + \mathcal{O}(1/\xi_{n2}^2)]
\end{eqnarray}
One has then
\begin{eqnarray}
\mathcal{L}_{\phi} \cong 2^{3/2}\lambda\bigg(\frac{\phi_{1}^2}{M_{5}^3}\bigg)^{1/2}\bigg(\frac{\beta}{M_5}\bigg)^{1/2}\tilde{m}e^{-\lambda_b\beta y_2}\sum_n q_n(x)\chi'^2
\end{eqnarray}
Thus since $\phi_{1}^2/M_{5}^3 \ll 1$ and $\beta/M_{5}^3 = \mathcal{O}(1)$, the coupling of the KK modes to the SM fields $\chi'$ is exponentially suppressed as long as $\lambda_b$ is not anomalously small, e.g. $\lambda_b \gtrsim 0.1$.

The final contribution to Eq.(103) comes from the ``radion'' mode
\begin{eqnarray}
Q_o = e^{2\beta y}[AJ_{\nu}(\xi_o) + BN_{\nu}(\xi_o)]
\end{eqnarray}
where $\xi_o(y) = (m_o/\beta)e^{\beta y}$, $m_o \cong \mu$ and $A$ and $B$ are given by Eqs.(76)-(78) with $f^T(y_1)$ and $f^T(y_2)$ related now by Eq.(112).  Since both $\xi_1$ and $\xi_2$ are small ($\xi_{\alpha} = (\mu/\beta)e^{\beta y_{\alpha}}$) the $f^T(y_1)$ parts of Eq.(76) are much smaller than the $f^T(y_2)$ parts.  Thus the ratio is ($\nu = 2 + \lambda_b$)
\begin{eqnarray}
\frac{N_{\nu}(\xi_2)}{e^{2\beta y_2}N_{\nu}(\xi_1)}\frac{\phi_o(y_1)}{\phi_o(y_2)}\frac{f^T(y_1)}{f^T(y_2)}\cong e^{-6\beta y_2}
\end{eqnarray}
Similarly, for the $B$ term one gets
\begin{eqnarray}
\bigg(\frac{\xi_2}{\xi_1}\bigg)^{\nu}\frac{\phi_o(y_1)f^T(y_1)}{e^{2\beta y_2}\phi_o(y_2)f^T(y_2)} = e^{-2(1-\lambda_b)\beta y_2}
\end{eqnarray}
Since as can be seen from Fig.2, the current tests of Newton's law requires $\lambda_b \lesssim 0.7$, again the $f^T(y_1)$ term is negligible.  We thus find
\begin{eqnarray}
A = \frac{1}{D}(\nu - 2 + \lambda_1)\phi_1N_{\nu}(\xi_1)\lambda_b(\lambda_b - \lambda_2)e^{2\beta y_2}f^T(y_2)
\end{eqnarray}
and using Eq.(84)
\begin{eqnarray}
A \cong \frac{\Gamma(1 + \nu)}{6}\frac{\lambda_b(\lambda_b - \lambda_2)}{\nu + 2 + \lambda_2}\bigg(\frac{\xi_1}{2}\bigg)^{-\nu}e^{-2\lambda_b\beta y_2}\phi_1f^T(y_2)
\end{eqnarray}
and similarly
\begin{eqnarray}
B \cong \frac{\pi}{6\Gamma(\nu)}\frac{\lambda_b(\lambda_b - \lambda_2)}{\nu + 2 + \lambda_2}\bigg(\frac{\xi_1}{2}\bigg)^{\nu}e^{-2\lambda_b\beta y_2}\phi_1f^T(y_2)
\end{eqnarray}
Since $\xi_o(y) \ll 1$ over the entire range of $y$, we can use the small $\xi_o$ terms of $J_{\nu}$ and $N_{\nu}$ in Eq.(147).  One finds then in coordinate space that
\begin{eqnarray}
Q_o(x^i,y) = Ce^{\nu\beta y}f^T(x^i,y_2)\,\,\,;\,\,\, C = \frac{\lambda_b(\lambda_b - \lambda_2)}{\nu + 2 + \lambda_2}\frac{\phi_1}{6}e^{2(1-\lambda_b)\beta y_2}
\end{eqnarray}
Thus $f^T(x^i,y_2)$ plays the role of the effective field for the $n = 0$ mode on the 4D brane at $y_2$.

We next determine the canonical normalization of $f^T(y_2)$.  From Eq.(122) the kinetic energy is
\begin{eqnarray}
\mathcal{L}_{KE}^{(o)} = -\frac{1}{2}\int d^4x \partial_if^T\eta^{ij}\partial_jf^T\int_{y_1}^{y_2}dy C^2e^{-2\beta y}e^{2\nu\beta y}
\end{eqnarray}
where we have neglected the non-diagonal terms between the $n=0$ and higher (KK) modes.  Hence writing
\begin{eqnarray}
q_o(x^i) = \mu_o f^T(x^i,y_2)
\end{eqnarray}
we have that
\begin{eqnarray}
\mu_{o}^2 = C^2\int_{y_1}^{y_2}dy e^{2(1 + \lambda_b)\beta y} \cong \frac{C^2}{2(1+\lambda_b)\beta}e^{2(1+\lambda_b)\beta y_2}
\end{eqnarray}
where $q_o(x^i)$ is the canonically normalized field.  Thus
\begin{eqnarray}
Q_o = [2(1+\lambda_b)\beta]^{1/2}e^{-(1+\lambda_b)\beta y_2}e^{\nu\beta y}q_o(x)
\end{eqnarray}
and the contribution of $Q_o$ to Eq.(103) is
\begin{eqnarray}
\mathcal{L}_{\phi}^{(o)} = 2\lambda(\frac{\phi_1}{M_{5}^3})^{1/2}[2(1 + \lambda_b)]^{1/2}\bigg(\frac{\beta}{M_{5}}\bigg)^{1/2}\frac{1}{\tilde{M}_5}q_o(x)[\tilde{m}^2\chi'^2]
\end{eqnarray}
where
\begin{eqnarray}
\tilde{M}_5 = M_5e^{-(1-\lambda_b)\beta y_2}
\end{eqnarray}
Thus the $q_o$ coupling is strongly suppressed by the large mass $\tilde{M}_5$ i.e. as $\lambda_b$ ranges from 0.1 to its maximum value (from the Newton law data in Fig.2) of $\lambda_b \cong 0.7$, $\tilde{M}_5$ ranges from $5\times10^{4}$GeV to $10^{14}$GeV.

We next calculate the mass of the normalized $q_o$ field.  From Eq.(122) the mass term is
\begin{eqnarray}
\mathcal{L}_{\phi}^{(o)m} = -\frac{1}{2}\int d^4x\int_{y_1}^{y_2}dye^{-4\beta y}[\gamma^2Q_{o}^2 + (\partial_5Q_o)^2]
\end{eqnarray}
and performing the $y$ integration gives
\begin{eqnarray}
\mathcal{L}_{\phi}^{(o)m} = -\frac{1}{2}\int d^4x[2\bigg(\frac{\nu - 1}{\nu - 2}\bigg)(\nu^2 - 2)\beta^2 e^{-2\beta y_2}]q_{o}^2(x)
\end{eqnarray}
Thus the $q_o$ mode has an effective mass of $\beta e^{-2\beta y_2} = \mathcal{O}$(TeV), which combined with Eq.(158) implies only a very small effect phenomenologically.

\section{Conclusions}
We have examined here the gravitational forces between point particles on the branes in the 5D Randall-Sundrum model with two branes and $S^1/Z_2$ symmetry (the RS1 model).  In terms of the orthogonal decomposition of the 4D part of the metric of Eq.(5), the static Newtonian forces should arise from $h_{00} = h_{00}^{TT} - f^T/3$ on the branes $y = y_{\alpha}, \alpha = 1,2$, where $h_{00}^{TT}$ is the transverse traceless part of the metric (and also contains the Kaluza-Klein corrections) and $f^T$ is the trace of the transverse part of the metric.  In order to impose the $S^1/Z_2$ boundary conditions correctly, it is necessary that the coordinate conditions chosen do not produce brane bending.  Thus we assume here only that $h_{5i}(x^i,y) = 0, i=0,1,2,3$.  While it is not possible to have $h_{55}$ vanish everywhere, one can assume it vanishes on the branes.  $h_{ij}^{TT}(x^i,y)$ is gauge invariant with respect to the remaining gauge freedom, and $f^T$ is gauge invariant on the branes (and plays the role of the radion).

Without a scalar field, the amplitude of $f^T(y_2)$ is enhanced by a factor $e^{2\beta y_2} \simeq 10^{32}$ making the theory  in serious disagreement with experiment\cite{ad}.  In this work we have included a scalar stabilizing field in the bulk $\phi(x^i,y) = \phi_{0}(y) + \delta\phi$, where $\phi_{0}(y)$ is the vacuum solution and $\delta\phi$ responds to matter on the branes.  The presence of $\phi$ can allow $f^T(y_2)$ to grow a mass, suppressing it.  To examine this possibility we considered the case where $\phi_{0}^2/M_{5}^3$ was small and $\phi_{0}(y)$ is a decreasing function of $y$.  Then one can obtain analytically the leading order corrections.  One finds that $f^T(y_2)$ does indeed grow a factor $e^{-\mu r}$ but is still enhanced by the $e^{2\beta y_2}$ factor.  Further, the mass $\mu$ of Eq.(91) is suppressed by the exponential factor $e^{-(1+\lambda_b)\beta y_2}$ where $\lambda_b$ defined in Eq.(74) is positive.  Thus whether the RS1 model is in agreement with 
 current small distance measurements of the Newtonian force law depends upon a subtle interplay between the amplitude enhancement and the exponential suppression of the mass.  Current data eliminates large parts of the parameter space.  The remaining allowed region is shown in Figs.1 and 2.

The Randall-Sundrum 1 model shows interesting features not intuitively expected.  Thus the fact that an exponential appears in the metric (a feature of the solution of the 5D vacuum Einstein equations) modifies the ideas of naturalness.  While one would expect that the mass of $f^T$ would scale by $\beta$, i.e. $\mu \sim \beta = \mathcal{O}(M_{Pl})$ (with perhaps a model dependent factor) the unexpected feature is the additional (model dependent) exponential factor, i.e. $\mu \sim \beta e^{-(1+\lambda_b)\beta y_2}$.  Since exponentials vary rapidly, they radically change the `natural' expectation of the size of $\mu$.  Such phenomena are intrinsic to the Randall-Sundrum model, since one is using exponentials to create an `unnatural' solution of the gauge hierarchy problem.  Further, the inverse of the very exponentials needed for the gauge hierarchy can enter from metric factors appearing in the denominator and do so in the amplitude of $f^T$ i.e. $f^T \sim e^{2\beta y_2}$.  It is thus remarkable that the theory can survive the experimental tests of the Newtonian force law.  Improvements of these experiments at distances smaller than $10\mu m$ will therefore further test the model.  Other tests of the model which could further reduce the allowed parameter space might occur when one introduces couplings of the scalar stabilizing field to Standard Model matter on the brane.  In Sec.8 we considered the simple example of a coupling proportional to $\phi^2\chi^2$ on the TeV brane (where $\chi$ is a scalar field, e.g. the Higgs field).  While a full discussion would require a numerical analysis, we find from an approximate analytic analysis that the effects of such a coupling are very small provided $\lambda_b$ is not anomalously small, i.e. $\lambda_b \gtrsim 0.1$ and obeys the Newton law data constraint $\lambda_b \leq 0.7$.

\section{Acknowledgement}
This work was supported in part by a National Science Foundation grant PHY-0101015.

\appendix
\section{Appendix}

\renewcommand{\theequation}{\Alph{section}.\arabic{equation}}
\setcounter{equation}{0}

In Sec.7 we considered a special solution for $A$ and $\phi_0$ of Eqs.(14-15).  We show here that this actually represents the leading terms of the more general Eqs.(11-13) when $\phi_{0}(x^i,y_1)$ is small (i.e. $\phi_{0}^2(y_1)/M_{5}^3 \ll 1$) and $\phi_0$ is a decreasing function of $y$.  

For the situation considered, we can expand $A'(y)$ and $V(\phi_0)$ in a power series in $\phi_{0}^2$
\begin{eqnarray}
A'(y) = \beta + \mathcal{O}(\phi_{0}^2) + \ldots\\
V(\phi_0) = \frac{1}{2}\gamma^2\phi_{0}^2 + \mathcal{O}(\phi_{0}^4) + \ldots
\end{eqnarray} 
where $\beta$ and $\gamma^2$ are arbitrary constants.  Eq.(13) in the bulk gives then
\begin{eqnarray}
\phi_{0}'' - 4\beta\phi_{0}' -\gamma^2\phi_{0} + \mathcal{O}(\phi_{0}^3) = 0
\end{eqnarray}
To leading order then (since $\phi_0$ is decreasing)
\begin{eqnarray}
\phi_{0} = \phi_{1}e^{-by} + \ldots\,\,\,;\,\,\, b > 0
\end{eqnarray}
where
\begin{eqnarray}
b^2 + 4\beta b -\gamma^2 = 0
\end{eqnarray}
It is convenient to introduce the parameter
\begin{eqnarray}
\lambda_b \equiv \frac{b}{\beta}
\end{eqnarray}
As discussed in \cite{dewolfe}, since $b>0$, the gauge hierarchy requires $\beta > 0$ so that
\begin{eqnarray}
\lambda_b > 0
\end{eqnarray}
and Eq.(A.5) implies
\begin{eqnarray}
\frac{\gamma^2}{\beta^2} = (\lambda_{b}^2 + 2)^2 - 4
\end{eqnarray}
Eqs.(11) and (12) imply in the bulk that
\begin{eqnarray}
3A'' = \frac{(\phi_{0}')^2}{M_{5}^3}
\end{eqnarray}
and inserting Eq.(A.4) (with the conventional boundary condition $A(0)=0$) one finds
\begin{eqnarray}
A = \beta y + \frac{\phi_{1}^2}{12M_{5}^3}e^{-2by} + \ldots
\end{eqnarray}
showing that the special solutions Eqs.(14) and (15) are the leading terms in the more general case.  The higher terms in Eqs.(A.4) and (A.10) are determined by the choice of bulk and brane potentials $V(\phi)$ and $V_{\alpha}(\phi)$.

As discussed in \cite{dewolfe}, it is still possible to achieve a solution of the gauge hierarchy when $b<0$ and $\phi_{0}(y)$ is an increasing function of $y$.  This situation is more complicated than the one treated in this paper since the terms $(\phi_{0}^2(y_1)/M_{5}^3)^2Q$ for example in Eq.(57) might become large and dominate.  Then an analytic solution as discussed here does not seem possible.

\section{Appendix}

\renewcommand{\theequation}{\Alph{section}.\arabic{equation}}
\setcounter{equation}{0}

We consider here the relation to some of the previous work in this area (Refs.\cite{csaki2,tanaka,tm,tk1,tk2}). 

In the analysis given here, we have considered gauge transformations that produce no brane bending (i.e. maintain the $S^1/Z_2$ boundary conditions).  Setting $h_{i5}$ to zero, the static gravitational force for particles on the branes is then governed by
\begin{eqnarray}
h_{00} = h_{00}^{TT} -\frac{1}{3}f^T
\end{eqnarray}
evaluated on the branes.  As seen from Eqs.(20) and (23), $h_{00}(y_{\alpha})$ is gauge invariant.  Thus the formalism describes the static gravitational energy in terms of gauge invariant parameters.  The usefulness of describing gravitational energies in gauge invariant language for 4D general relativity is well known (e.g. Refs.\cite{adm,adm2}) and we make use of these ideas in the RS1 5D model.  We first consider Ref.\cite{csaki2} which assumes a metric of the form
\begin{eqnarray}
ds^2 = e^{-2A}[\eta_{ij} - 2F\eta_{ij}]dx^idx^j + (1+2G)dy^2
\end{eqnarray}
which implies in the notation of Sec.2
\begin{eqnarray}
h_{55}\equiv 2G\,\,\,;\,\,\,;f^T\equiv -6F
\end{eqnarray}
but in addition from the general form of Eq.(6)
\begin{eqnarray}
h_{ij}^{TT} = 0\\h_{i}^T = 0\\\Box^2 h^L = \frac{1}{3}f^T
\end{eqnarray}
Eq.(B.4) implies that the ``TT'' modes are not being considered.  (They do indeed contribute to the gravitational forces as seen in Eq.(65).)  Eq.(B.5) is not a priori obvious, but as seen from Eqs.(41,42), the $R_{j5}$ equation combined with a gauge choice allows one to set $h_{i}^T$ to zero.

The problematical choice is Eq.(B.6).  Assuming Eq.(B.6), Ref.\cite{csaki2} argues that the $\partial_i\partial_j$ part of the $R_{ij}$ equation allows one further to set $G=2F$ i.e. $h_{55} = -2f^T/3$.  We find in fact from the $R_{ij}$ equation that
\begin{eqnarray}
\Box^{-2}e^{-2A}[\frac{1}{2}\partial_{5}^2 - 2A'\partial_5](\Box^2h^L - \frac{1}{3}f^T) + \frac{1}{2}(h_{55} + \frac{2}{3}f^T) = 0
\end{eqnarray}
Thus if Eq.(B.6) holds then indeed $h_{55} = -2f^T/3$.  Thus one must check whether Eq.(B.6) can be achieved by a gauge choice.  From Eqs.(23) and (25) on can easily see that starting in an arbitrary gauge one can in fact achieve Eq.(B.6) by the gauge transformation
\begin{eqnarray}
2e^{2A}\Box^2\xi^5 = \partial_5(\Box^2h^L -\frac{1}{3}f^T)
\end{eqnarray}
where the right hand side is in the arbitrary gauge.  However, the Einstein equations imply Eq.(36) so that Eq.(20) is violated when $T(y_{\alpha})\neq 0$, i.e. matter is on the branes.  Eq.(B.6) then implies a gauge with brane bending and hence with non-trivial boundary conditions to be imposed.  Actually Ref.\cite{csaki2} does the analysis of the radion mass $m_r$ under the assumption that there is no matter on the branes, so that the analysis is actually valid.  It should be noted that the radion ($h_{55}$) is not gauge invariant and that while (in the stiff potential limit) $m_r$ does equal the mass $\mu$ of Eq.(91), the former refers to the radion mass in the bulk while the latter refers to the short range part of the gauge invariant gravitational potential of matter on the branes.  However with the actual gauge chosen, $h_{55}=-2f^T/3$ and the two functions are directly related on the branes.

Ref.\cite{csaki2} does subsequently put Standard Model (SM) matter on the branes but still imposes Eq.(B.6) which then is inconsistent with the boundary conditions.  As can be seen from Sec.2, the coupling of the 5D fields to the SM on the branes actually involves five fields ( $h_{ij}^{TT}$, $h_{55}$, $f^T$, $h^L$, and $\delta\phi$) of which one may be eliminated by Eq.(40).  The remaining coupled set should then be treated quantum mechanically and subject to the matter boundary conditions.  Thus the quantum field treatment of SM coupling is much more complicated than having only a single bulk field F (as in Ref.\cite{csaki2}).

In Ref.\cite{tm} the gauge choice of Eqs.(B.5) and (B.6) was also used (referred to there as the ``Newton gauge'') but these authors do include matter on the branes.  They state correctly that this gauge implies brane bending.  To impose the boundary conditions, Ref.\cite{tm} makes a transformation to 'locally' Gaussian normal coordinates where $h_{55}$ vanishes in the vacinity of the branes.  This requires a different coordinate frame for each brane.  The authors assume the local Gaussian coordinates do not have brane bending and impose normal boundary conditions there.  (A derivation that local Gaussian coordinates are in fact free of brane bending was given in \cite{ad}.  It was also shown there that global Gaussian coordinates will generally possess brane bending on one or both branes, and so the assumption that this is not the case in Ref.\cite{tanaka} is not generally valid.)  Thus in principle, the formalism of Ref.\cite{tm} should be able to deduce the gravitational f
 orces between particles.  They do in fact calculate the radion mass (in the stiff potential limit) and show that the large $r$ limit recovers the long range Newtonian force.  However they do not generate the short range corrections to the Newtonian force of Eqs.(88) and (89) due to the fact that they make an expansion in powers of $\Box^2/m_{r}^2$ (which actually sum to the short range corrections) and argue that these correction terms are unobservable.  They also do not discuss the important exponential enhancement of the short range force appearing in Eq.(88) which is what makes these short range corrections within striking distance of experiment (as seen in Fig.1).


\end{document}